\documentclass{ws-ijmpd}
\usepackage[super,compress]{cite}
\usepackage{siunitx}
\usepackage{url}
\usepackage{nicefrac}
\begin{document}

\markboth{C. R. Farrugia \& J. Sultana}
{The Effects of Spatial Curvature on Cosmic Evolution}

\catchline{}{}{}{}{}
%
%%%%%%%%%%%%%%%%%%%%%%%%%%%%%%%%%%%%%%%%%%%%%%%%%%%%%%%%%%%%%%%%%%%%

\title{THE EFFECTS OF SPATIAL CURVATURE ON COSMIC EVOLUTION}

\author{CHRISTINE R. FARRUGIA}

\address{Department of Mathematics, University of Malta,\\
Msida, MSD 2080,
Malta\\
christine.farrugia.08@um.edu.mt}

\author{JOSEPH SULTANA}

\address{Department of Mathematics, University of Malta,\\
Msida, MSD 2080, Malta\\
joseph.sultana@um.edu.mt}

\maketitle

\begin{history}
\received{Day Month Year}
\revised{Day Month Year}
\end{history}

\begin{abstract}
As evidenced by a great number of works, it is common practice to assume that the Universe is flat. However, the majority of studies which make use of observational data to constrain the curvature density parameter are premised on the $\Lambda$CDM cosmology, or extensions thereof. On the other hand, fitting the data to models with a time-varying dark energy equation of state can, in some cases, accommodate a non-flat Universe. Several authors caution that if the assumption of spatial flatness is wrong, it could veer any efforts to construct a dark energy model completely off course, even if the curvature is in reality very small. We thus consider a number of alternative dynamical dark energy models that represent the complete cosmological scenario, and investigate the effects of spatial curvature on the evolution. We find that for a closed Universe, the transition to the epoch of decelerated expansion would be delayed with respect to the flat case. So would the start of the current dark energy-dominated era. This would be accompanied by a larger inflationary acceleration, as well as a larger subsequent deceleration. The opposite behavior is observed if the Universe is open.  
\end{abstract}


\ccode{PACS numbers:}

%\tableofcontents

\section{Introduction}	

The Universe is commonly modeled with zero spatial curvature. This is often based on the consistency of a flat Universe with a number of studies which have used data from sources such as type Ia supernovae (SNeIa), Baryon Acoustic Oscillations (BAOs) and the Cosmic Microwave Background (CMB) power spectrum to constrain several cosmological parameters, including the curvature density parameter $\Omega_k$ (see, for instance,\footnote{In Ref.~\refcite{WMAP}, the authors find that $|\Omega_k|<\num{0.0094}$ (at 95\% confidence) when they combine the data from the WMAP experiment and other sources. They however remark that `a small deviation from flatness is expected and is worthy of future searches'.} Refs.~\citen{WMAP, MAXIMA, Boomerang, Boomerang2, Planck, Tegmark2004, Eisenstein, Tegmark, Seljak, Wang, Dodelson, Tegmark2001}). However, these studies usually fit the data to the $\Lambda$CDM model (or extensions thereof), arguably the most popular cosmological model today. In $\Lambda$CDM, the elusive \emph{dark energy} -- a form of energy whose gravitational repulsion is thought to be behind the present cosmic acceleration -- is modeled with equation of state (EoS) $\rho_\Lambda = -p_\Lambda = \Lambda c^4/8\pi G$. Thus the dark energy density $\rho_\Lambda$, being constant, starts to dominate at late times, because the densities of matter and radiation get diluted as the Universe expands.

Despite its numerous successes, $\Lambda$CDM is not without its limitations. Prominent among these is the \emph{cosmological constant} or \emph{smallness} problem, so called because the theoretical value of $\Lambda$ is larger than the observed value by a factor of $\sim\num{e120}$. Another difficulty is encountered when one considers that the densities of matter and dark energy appear to have the same order of magnitude at present. In the $\Lambda$CDM cosmology, dark energy has a constant density, but this is not so in the case of matter. Consequently, the explanation that emerges from $\Lambda$CDM is that we seem to be living at just the right time for the two densities to be comparable. This is, however, rather implausible, and constitutes a shortcoming which we call \emph{the coincidence problem}. 

Such inconsistencies have led to the emergence of a number of different cosmological models. Most of these fall into either of two categories. \emph{Modified gravity models} focus on providing an explanation for the effects attributed to dark energy by describing the geometry of the spacetime manifold in a way alternative to General Relativity. They include $f(R)$ Gravity, Weyl Gravity, and Scalar--Tensor theory, to mention a few. On the other hand, in \emph{modified matter models}, the energy-momentum tensor is altered to incorporate a component that provides a negative pressure \cite{Yoo} and can hence drive the late-time cosmic acceleration. Prominent examples are the Quintessence, K-essence and Phantom (or Ghost) models \cite{Miao}. The practice of considering the dark energy EoS to be dynamical, choosing a suitable parametrization and then attempting to constrain the parameters is also popular in the literature.\cite{Gong, Ichikawa2006, Zhao, Ichikawa}

In most modified matter models, the cosmological constant is replaced by a dark energy density $\rho_{\text{d}}$ that is a (direct or indirect) function of cosmic time. In light of this, they are often referred to as \emph{dynamical dark energy models}. The question naturally arises whether it is justified to extend the assumption of flatness to such models. To be fair, a number of studies have obtained constraints on the spatial curvature in several modified matter scenarios, and they have found a flat geometry to be favored.\cite{Gong, Ichikawa2006, Zhao, BonillaRivera} Other studies, however, have shown that when cosmological data is fitted to models with a time-varying dark energy EoS, the constraints on $\Omega_k$ may accommodate a non-flat Universe.\cite{Ichikawa, IchikawaK} It is important to bear in mind that observational bounds on dark energy and curvature may be mutually dependent,\cite{Wang2007} and degeneracies often set in when one tries to constrain the EoS of the former simultaneously with the latter.\cite{Virey} Indeed, if the density of dark energy is allowed to vary freely with time, constraints on the geometry of the Universe may not only become less stringent, but may even depend on the properties that dark energy is assumed to have at early times.\cite{Wang2007} The authors of Ref.~\refcite{Virey} report that models with curvature and dynamical dark energy can actually point to a flat $\Lambda$CDM cosmology when the fit does not take a potential dark energy evolution into account. It was also shown that a closed Universe consisting of dust and quintessence (a form of dark energy represented by a scalar field) may mimic a flat, accelerating $\Lambda$CDM Universe at late times.\cite{delCampo} Furthermore, it seems possible that recent SNeIa data is in tension with the flat $\Lambda$CDM hypothesis,\cite{Rest} or may actually favor a non-flat Universe.\cite{Kumar, Liddle, Delubac} 

The assumption of spatial flatness is sometimes also justified by invoking the prediction made by many inflationary models: namely, that the rapid inflationary acceleration would have `smoothed out' any initial curvature. However, as pointed out in Ref.~\refcite{Ichikawa}, a better approach would be to look for any potential curvature as a means of testing the inflationary paradigm, rather than the other way round. Additionally, the literature contains several examples of inflationary models that do not necessarily give rise to a Universe with a Euclidean spatial geometry.\cite{Bucher, Linde}

Despite all this, the majority of works put forward continue to be based on the assumption that the Universe is flat. In this paper, therefore, we investigate the  consequences of introducing curvature into different cosmological models alternative to $\Lambda$CDM. We consider models that provide a complete description of the expansion history of the Universe, starting with an early inflationary phase which then evolves into an epoch of deceleration; this is in turn followed by the present period of dark energy-driven acceleration. In particular, we adopt Kremer's model,\cite{Kremer} in which the Universe is modeled as a dissipative mixture of a Van der Waals (VdW) fluid and a dark energy component represented by either Quintessence or the Chaplygin gas. In Section 2, we extend Kremer's model by including spatial curvature and investigating its effects on the entire evolution of the Universe. Then we consider two cosmological models with a dark energy component represented by a dynamical vacuum term $\Lambda(t)$, and with a matter distribution in the form of either a VdW fluid (Section 2) or with the customary direct proportionality between the energy density and pressure (Section 3). In each case, we again note the effects of curvature on the evolution. Results and conclusions are presented in Section 4. 

Throughout this work, we set $c = 8\pi G/3 = 1$. All numerical analysis was carried out using $\textsc{Wolfram Mathematica}^{\text{\textregistered}} 10$. The figures were created using the \textsc{LevelScheme} scientific figure preparation system.\cite{LevelScheme}

\section{Matter as a Van der Waals Fluid}
We first consider the Universe as a matter distribution with the VdW EoS and do not include any dark energy. Building on Kremer's model,\cite{Kremer} we repeat his analysis but take the possibility of non-zero spatial curvature into account. The starting point is the following Friedmann equation:
\begin{equation}
H^2=\rho_\text{m}-\frac{\kappa}{a^2}
\label{1stFriedmann}
\end{equation}
where $H$ is the Hubble parameter and $\rho_\text{m}$ the energy density of matter. The scale factor $R$ is normalized with respect to $R_*$, the value it takes when $H = 1$ (we refer to the time at which this happens as $t_*$. In other words, $H(t_*) = H_* = 1$). Thus, $a$ is defined as the ratio $R/R_*$. We then shift the time scale by putting $\tilde{t}=t-t_*$, where $t$ is the customary cosmic time. The Hubble parameter $H$ is equal to $\dot{a}/a$ and the dot denotes differentiation with respect to $\tilde{t}$. At $\tilde{t}=0$, we have that $a(0)=\dot{a}(0)=1$.

As for the spatial curvature $\kappa$, this can be expressed in terms of the normalized spatial curvature $k$ ($k = 0$ or $\pm 1$) as $\kappa = k/R_*^2$. Hence, $\kappa$ has dimensions of $(\text{length})^{-2}$ and can take any real value.\cite{Carroll}

Differentiating Eq. ($\ref{1stFriedmann}$) with respect to $\tilde{t}$ and using the energy conservation equation:
\begin{equation}
\dot{\rho}_\text{m}=-3H(\rho_\text{m}+p_\text{m})
\label{energycons}
\end{equation}
to substitute for $\dot{\rho}_\text{m}$, one gets:
\begin{equation}
3 a^2(\rho_\text{m}+p_\text{m})-2(\kappa+\dot{a}^2)+2a\ddot{a}=0.
\label{intermediate}
\end{equation}  
The relation between the pressure $p_\text{m}$ of the matter content and its energy density is expressed as the VdW EoS:\cite{Johnston}
\begin{equation}
p_\text{m}=\left[\frac{8w\phantom{i}\nicefrac{\rho_\text{m}}{\rho_\text{m}^\text{c}}}{3-\nicefrac{\rho_\text{m}}{\rho_\text{m}^\text{c}}}-3\left(\frac{\rho_\text{m}}{\rho_\text{m}^\text{c}}\right)^2\right]p_\text{m}^\text{c}.
\label{VdWEoS1}
\end{equation}
In the original classical thermodynamics context, $w$ is a dimensionless parameter equivalent to the ratio of the fluid temperature to its critical value, whereas $\rho_\text{m}^\text{c}$ and $p_\text{m}^\text{c}$ are the critical energy density and pressure of the fluid, respectively. In a plot of pressure against volume, the critical point of a VdW fluid would correspond to the vertex of the region in which the gas and liquid phases coexist.\cite{Callen} 

When Eq. ($\ref{VdWEoS1}$) is used in a cosmological setting, $w$ is often identified with the constant of proportionality in the conventional barotropic formula $p_\text{m}=w\rho_\text{m}$, because this is essentially the form that ($\ref{VdWEoS1}$) takes for small values of $\rho_\text{m}$.\cite{Kremer2004} It is also common practice to reduce the number of free parameters by setting $\rho_\text{m}^\text{c} = p_\text{m}^\text{c} = 1$. The VdW EoS can then be rewritten concisely as:\cite{Kremer2004}
\begin{equation}
p_\text{m}=\frac{8w\rho_\text{m}}{3-\rho_\text{m}}-3\rho_\text{m}^2.
\label{VdWEoS}
\end{equation}
Replacing the pressure $p_\text{m}$ in ($\ref{intermediate}$) with ($\ref{VdWEoS}$), and $\rho_\text{m}$ with the corresponding expression from ($\ref{1stFriedmann}$), we obtain the following:
\begin{equation}
\kappa+\dot{a}^2-\frac{9(\kappa+\dot{a}^2)^2}{a^2}-\frac{24 w a^2(\kappa+\dot{a}^2)}{\kappa-3 a^2+\dot{a}^2}+2a\ddot{a}=0.
\label{MainEq}
\end{equation}
When $\kappa=0$, ($\ref{MainEq}$) reduces to Kremer's\cite{Kremer} Eq. (5), which is however expressed as a first-order differential equation in $H$.
\begin{figure}[b]
\centering
\includegraphics[width=10cm]{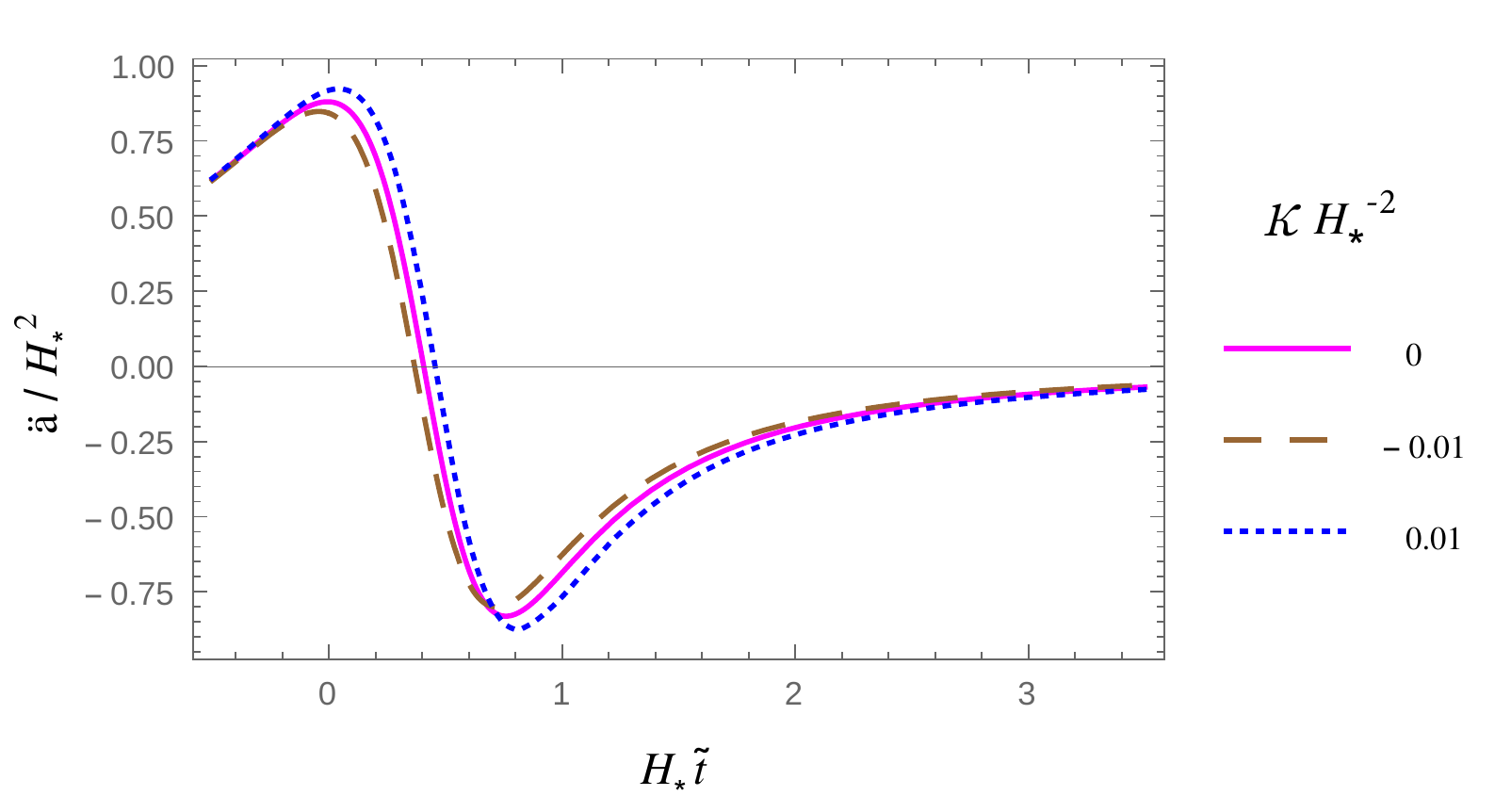}
\caption{\label{VdWacc}The variation of acceleration with time for a Universe composed of a VdW fluid.}
\end{figure}
\begin{figure}[b]
\centering
\includegraphics[width=10cm]{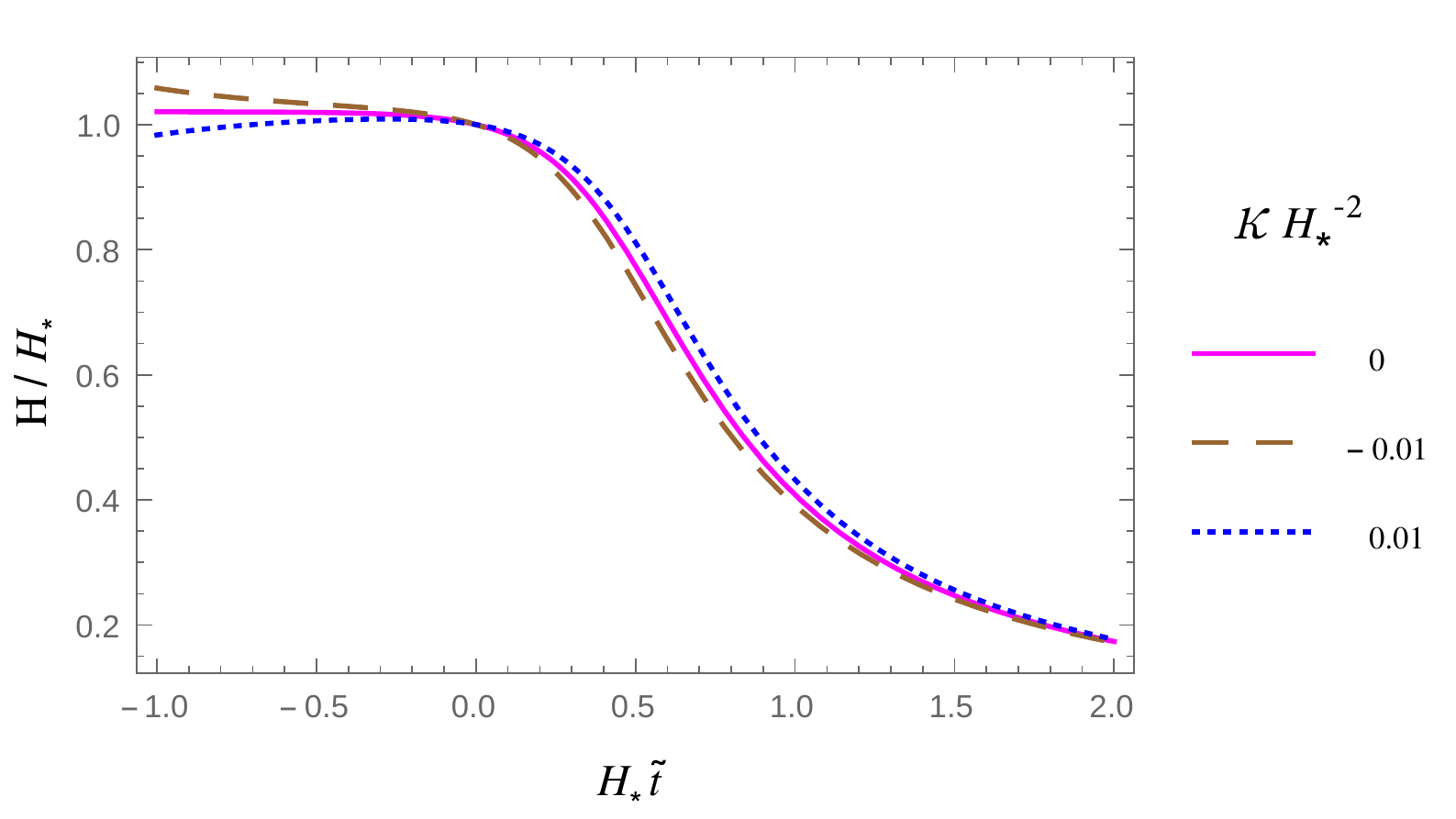}
\caption{\label{VdWH}The variation of the Hubble parameter with time for a Universe composed of a VdW fluid.}
\end{figure}

Eq. ($\ref{MainEq}$) can be solved numerically by choosing an appropriate set of initial conditions. As described above, we have that $a(0) = \dot{a}(0) = 1$, in line with Kremer's work. The parameter $w$ is fixed at \num{0.52}, while $\kappa$ is assigned the values zero (denoting a flat Universe, i.e. one endowed with Euclidean spatial geometry\cite{Lasenby}), \num{0.01} and \num{-0.01}. A positive $\kappa$ denotes a \emph{closed} Universe. If embedded in four-dimensional Euclidean space, the spatial part of such a Universe would have the shape of a three-dimensional sphere, whose volume is finite.\cite{Lasenby} On the other hand, negative $\kappa$ implies an \emph{open} Universe. In this case, the spatial section is more commonly embedded in four-dimensional Minkowski space, where it takes the form of a three-dimensional hyperboloid and consequently has unbounded volume.\cite{Lasenby} Such embedding is merely a means of elucidating the meaning of spatial curvature; one should otherwise keep in mind that the Universe is an entity in itself and avoid thinking of it as an embedding in a higher-dimensional space. 

\begin{figure}[b]
\centering
\includegraphics[width=10cm]{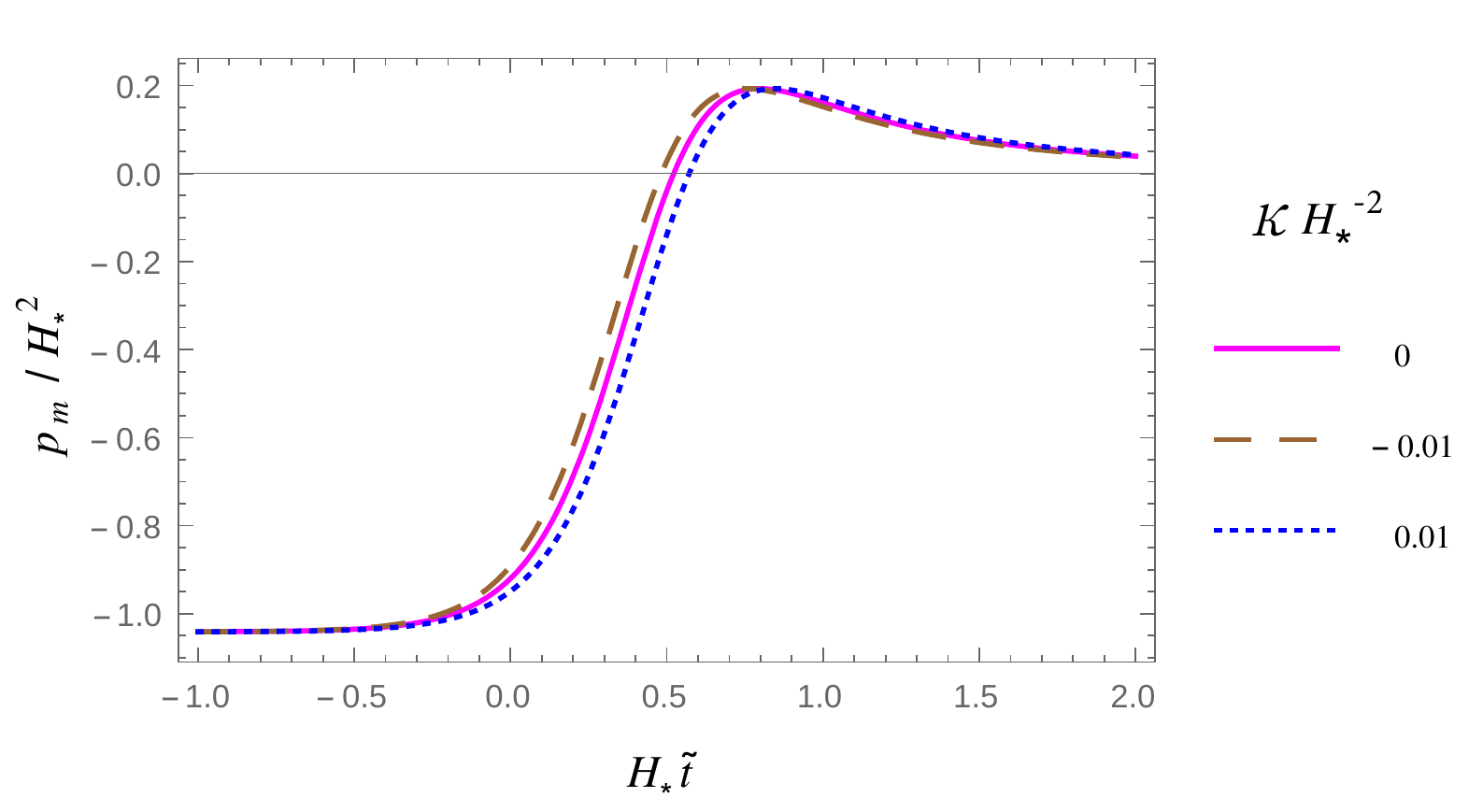}
\caption{\label{VdWp}The variation of pressure with time for a Universe composed of a VdW fluid.}
\end{figure}

\begin{figure}[b]
\centering
\includegraphics[width=10cm]{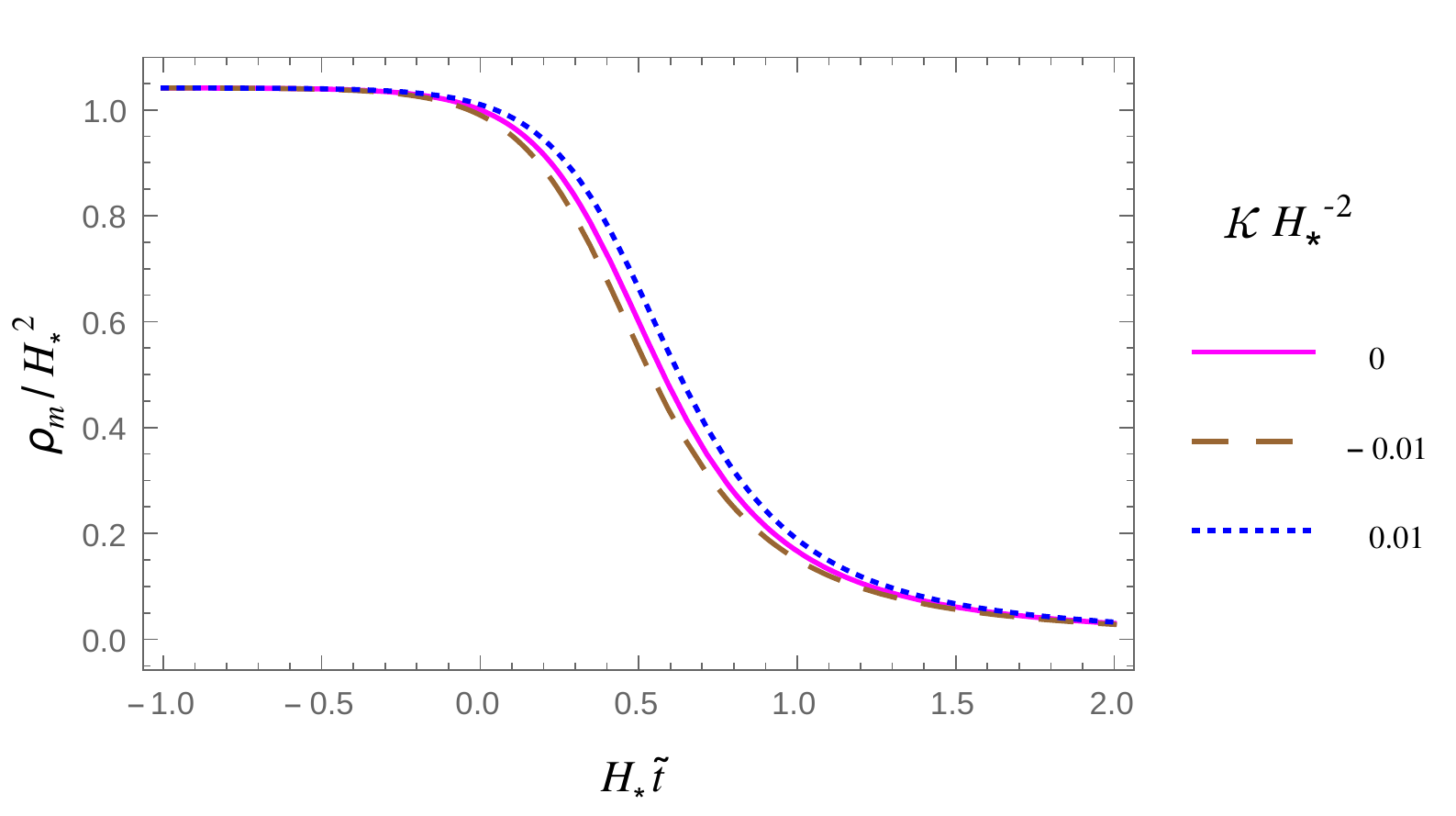}
\caption{\label{VdWrho}The variation of energy density with time for a Universe composed of a VdW fluid.}
\end{figure}

The resulting plots are shown in figures $\ref{VdWacc}$, $\ref{VdWH}$, $\ref{VdWp}$ and $\ref{VdWrho}$. 

The most widely-accepted view nowadays with regards to the different stages of cosmic evolution is that, shortly after the Big Bang, the Universe underwent a period of accelerated expansion known as \emph{inflation},\cite{Guth, Linde, Albrecht} and that a graceful exit then led to a radiation-dominated epoch (and a later matter-dominated one) during which deceleration occurred.\cite{Turner2002, Shapiro} This was followed by further acceleration\cite{Riess, Perlmutter, Gong2006, Caldwell2009} -- which extends to the present time and is thought to be one of the most notable sources of evidence for dark energy. As pointed out by Kremer, modeling the Universe as a VdW fluid can correctly describe the transition from the inflationary epoch to the subsequent deceleration, but fails to yield the proposed late-time acceleration (Fig. $\ref{VdWacc}$). 

The evolution of the Universe is determined by the pressure of the cosmic fluid. In fact, Fig. $\ref{VdWp}$ shows that when this fluid is described by the VdW EoS (Eq. ($\ref{VdWEoS}$)), $p_\text{m}$ would initially be negative (resulting in inflation -- Fig. $\ref{VdWacc}$). It would then start to increase and become positive (causing the acceleration to decrease and deceleration to set in), finally decreasing asymptotically to zero (with deceleration following suit) rather than becoming negative again (which would cause the Universe to accelerate).
  
Another point of interest is the way the presence of curvature affects the cosmic evolution. From Fig. $\ref{VdWacc}$, it can be seen that positive curvature delays the onset of deceleration (with respect to the flat case), while negative curvature causes it to occur earlier. The maximum value of the acceleration/deceleration also changes with $\kappa$: it is largest for the closed Universe, and least for the open one. Furthermore, small variations in the parameters or the initial conditions only alter the magnitude of these effects, and not the behavior of the evolution. The closed Universe would still be characterized by the largest acceleration and deceleration, and the transition from one to the other would still occur later than it does for the flat and open geometries.     

\subsection{Dark energy as Quintessence or a Chaplygin gas}
To alleviate the problem of the absence of late-time acceleration, Kremer\cite{Kremer} proposes adding a dark energy component to the VdW matter distribution and taking dissipative effects into account. Thus Eq. ($\ref{1stFriedmann}$) becomes
\begin{equation}
H^2=\rho_\text{m}+\rho_\text{d}-\frac{\kappa}{a^2}
\label{1stFriedmann2}
\end{equation}
where $\rho_\text{d}$ is the dark energy density. Energy conservation can now be expressed as two decoupled equations: 
\begin{align}
\dot{\rho}_\text{d}&=-3H(\rho_\text{d}+p_\text{d});
\label{energycons2}\\[0.5em]
\dot{\rho}_\text{m}&=-3H(\rho_\text{m}+p_\text{m}+\varpi)
\label{energycons3}
\end{align}
since we assume that matter and dark energy are non-interacting.\cite{Kremer} In Eq. ($\ref{energycons2}$), $p_\text{d}$ represents the pressure associated with dark energy. The term $-3H\varpi$ on the right-hand side of ($\ref{energycons3}$), where $\varpi$ is the non-equilibrium pressure, describes the rate at which energy (density) is transferred irreversibly from the gravitational field to the VdW fluid. Indeed, the possibility of expressing energy conservation as two distinct equations also reflects the fact that dark energy is considered to be only minimally coupled to the gravitational field \cite{Kremer}.

In second-order non-equilibrium thermodynamics, the evolution of $\varpi$ is described by the equation:
\begin{equation}
\tau\dot{\varpi}+\varpi=-3\eta H.
\label{NonEqPressure2}
\end{equation}
Here, $\tau$ is the characteristic time and $\eta$ the coefficient of bulk viscosity.\cite{Kremer} A derivation of this equation\footnote{Kremer\cite{Kremer} cites Ref.~\refcite{KremerBook}.} can be found in Ref.~\refcite{Chakraborty}, where the authors make use of the seminal work of Israel and Stewart.\cite{Israel} As in Ref.~\refcite{Kremer}, we take $\eta$ to be directly proportional to the energy density\cite{Murphy} of the mixture ($\eta = \alpha \rho_T$, where $\rho_T = \rho_\text{m}+\rho_\text{d}$ and $\alpha$ is a constant) and set $\tau$ equal to\cite{Belinskii} $\eta/\rho_T$. Hence $\tau = \alpha$, and Eq. ($\ref{NonEqPressure2}$) becomes $\alpha\dot{\varpi}+\varpi=-3(\alpha \rho_T) H$. Eq. ($\ref{1stFriedmann2}$) can then be utilized to eliminate $\rho_T$, getting:
\begin{equation}
\dot{\varpi}+\frac{1}{\alpha}\varpi = -3 H\left(H^2+\frac{\kappa}{a^2}\right). 
\label{NonEqPressure}
\end{equation}
The hydrostatic pressure $p_\text{m}$ and energy density $\rho_\text{m}$ are again related by the VdW EoS ($\ref{VdWEoS}$). As for the dark energy EoS, we follow Kremer's work once more and model dark energy as either Quintessence or a Chaplygin gas.\cite{Kremer} A similar model consisting of Quintessence and a viscous VdW fluid in a Lyra manifold is considered in Ref.~\refcite{Khurshudyan}, although the authors also include a constant or dynamical $\Lambda$ and restrict their analysis to a flat geometry. Furthermore, they consider dark energy and dark matter to be interacting, which is not the case in our work. 

\subsubsection{Quintessence}
In this case, the dark energy EoS reads\cite{Kremer}
\begin{equation}
p_\text{d}=w_\text{d} \rho_\text{d};~~~w_\text{d}<-1/3
\label{QuintessenceEoS}
\end{equation}
where $w_\text{d}$ is a constant. When inserted into ($\ref{energycons2}$), Eq. ($\ref{QuintessenceEoS}$) allows us to solve for $\rho_\text{d}$, getting
\begin{equation}
\rho_\text{d}=\rho_\text{d}^0~a^{-3(1+w_\text{d})}
\label{darkenergydensity}
\end{equation} 
with $\rho_\text{d}^0$ the dark energy density at $\tilde{t}=0$. We then determine the time derivative of ($\ref{1stFriedmann2}$), and into this we insert equations ($\ref{energycons2}$) and ($\ref{energycons3}$), followed by ($\ref{VdWEoS}$) and ($\ref{QuintessenceEoS}$). Finally, the VdW energy density, $\rho_\text{m}$, is replaced with the corresponding expression from Eq. ($\ref{1stFriedmann2}$), while ($\ref{darkenergydensity}$) is substituted for $\rho_\text{d}$. The resulting equation -- together with ($\ref{NonEqPressure}$) -- determines the evolution of a Universe filled with a VdW matter distribution and Quintessence, and is given by:
\begin{align}
&3f(\tilde{t},\kappa)\left[\frac{8w a^2}{3a^2+\rho_\text{d}^0 a^{-(1+3w_\text{d})}-(\kappa+\dot{a}^2)}-\frac{3}{a^2}f(\tilde{t},\kappa)\right]+\frac{3w_\text{d} \rho_\text{d}^0}{a^{1+3w_\text{d}}}+\kappa+3a^2\varpi+\dot{a}^2\notag\\+~&2a \ddot{a}=0
\label{MainEq2}
\end{align}
where 
\begin{equation}
f(\tilde{t},\kappa)=-\frac{\rho_\text{d}^0}{ a^{1+3w_\text{d}}}+\kappa+\dot{a}^2.
\end{equation}
When $\kappa=0$ (and with some modifications), Eq.($\ref{MainEq2}$) reduces to Kremer's\cite{Kremer} Eq. (13), although the latter is expressed as a first-order differential equation in $H$.
\begin{figure}[b]
\centering
\includegraphics[width=10cm]{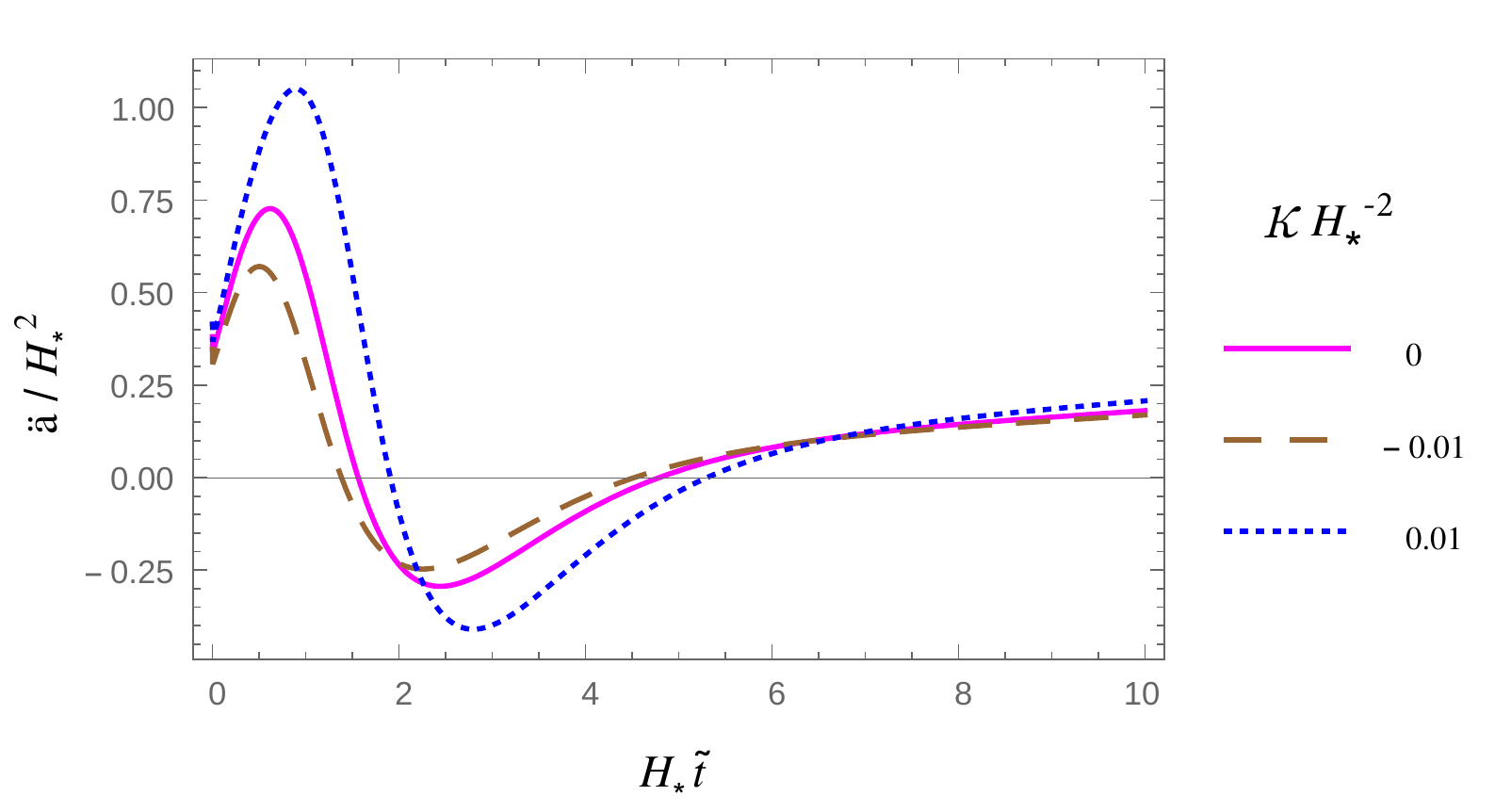}
\caption{\label{Quintacc}The variation of acceleration with time for a Universe composed of a VdW matter distribution and Quintessence.}
\end{figure}
\begin{figure}[tb]
\centering
\includegraphics[width=10cm]{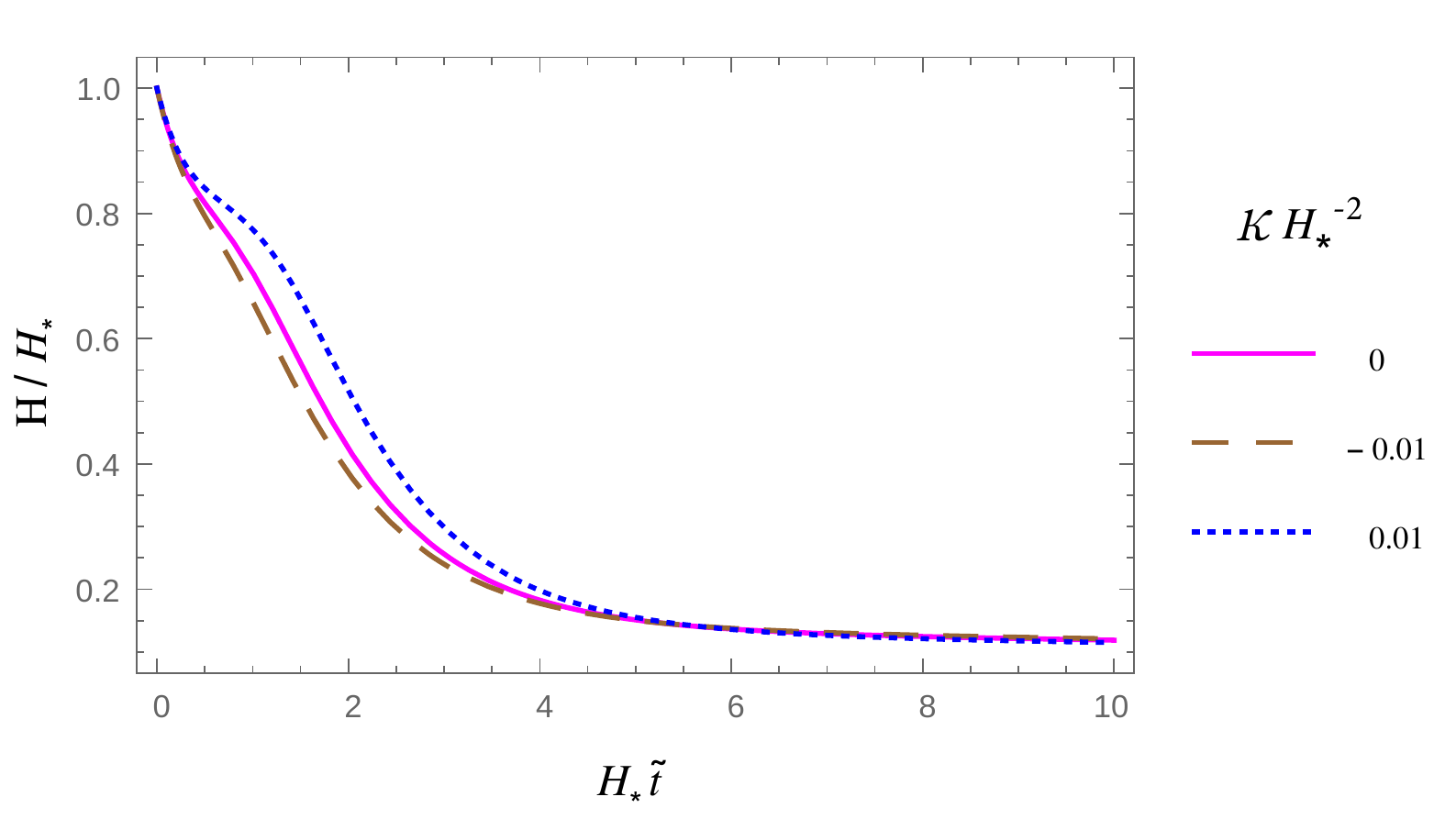}
\caption{\label{QuintH}The variation of the Hubble parameter with time for a Universe composed of a VdW matter distribution and Quintessence.}
\end{figure}

\begin{figure}
\centering
\includegraphics[width=10cm]{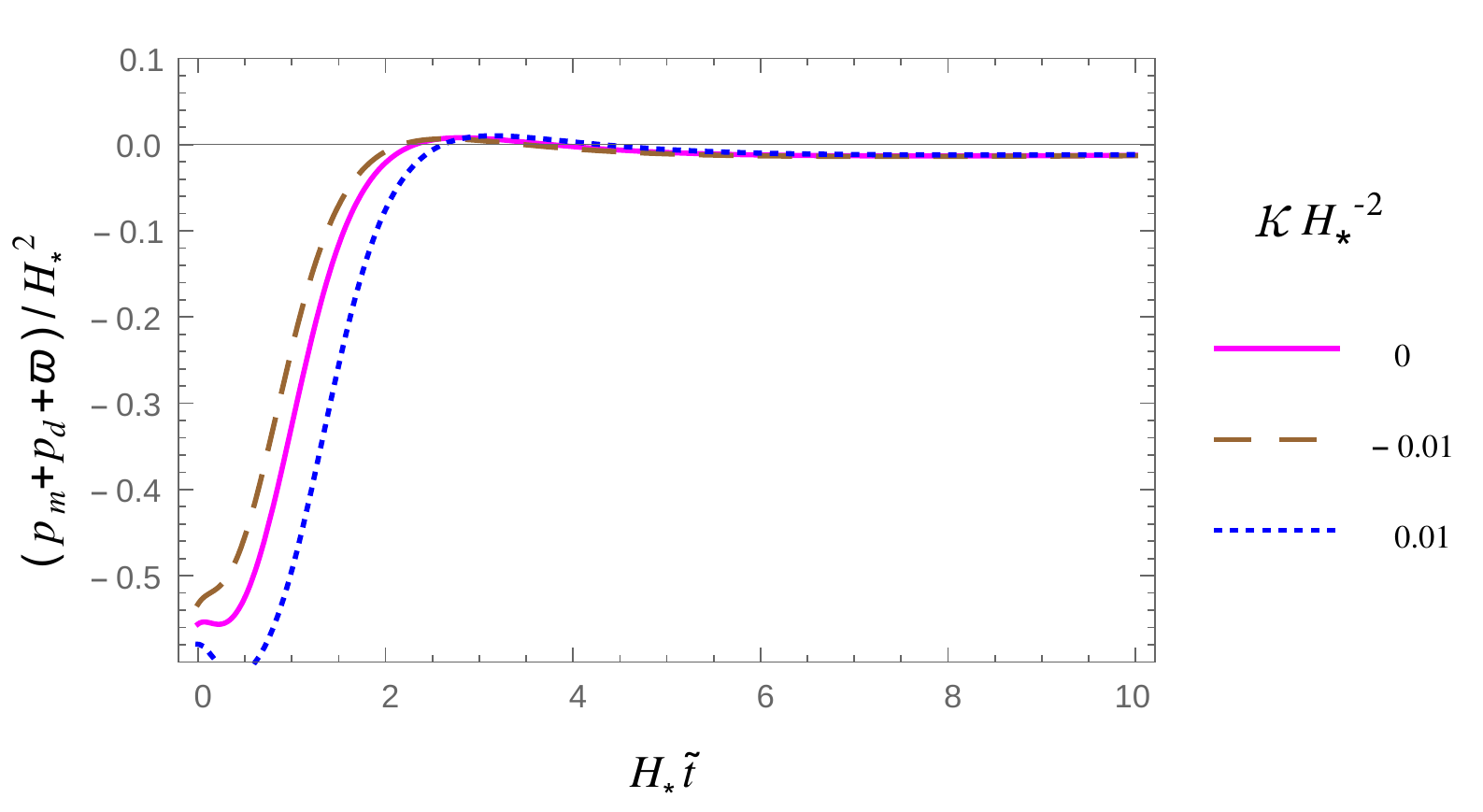}
\begin{picture}(0,0)
\put(-172.5,38){\includegraphics[height=2.5cm]{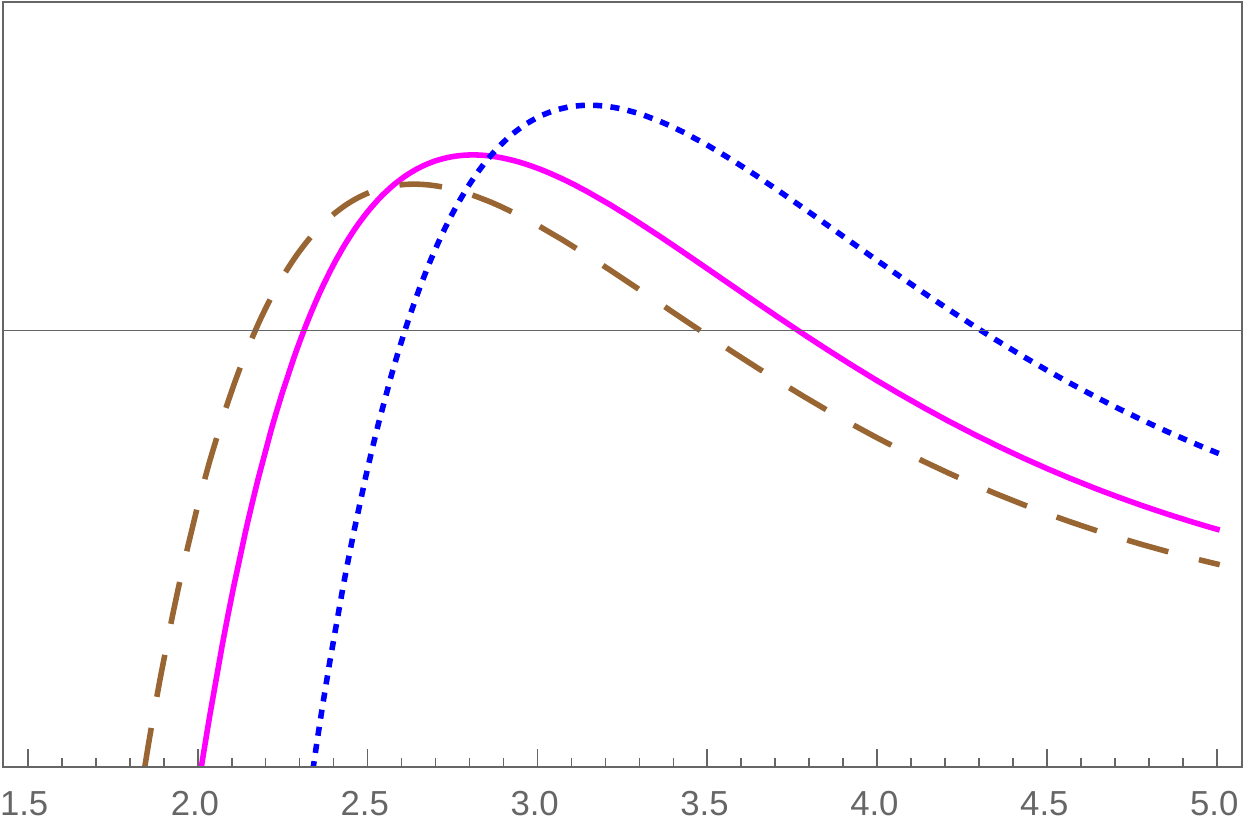}}
\end{picture}
\caption{\label{Quintp}The variation of the total pressure with time for a Universe composed of a VdW matter distribution and Quintessence. The peaks of the curves are shown at greater resolution (inset).}
\end{figure}

\begin{figure}
\centering
\includegraphics[width=10cm]{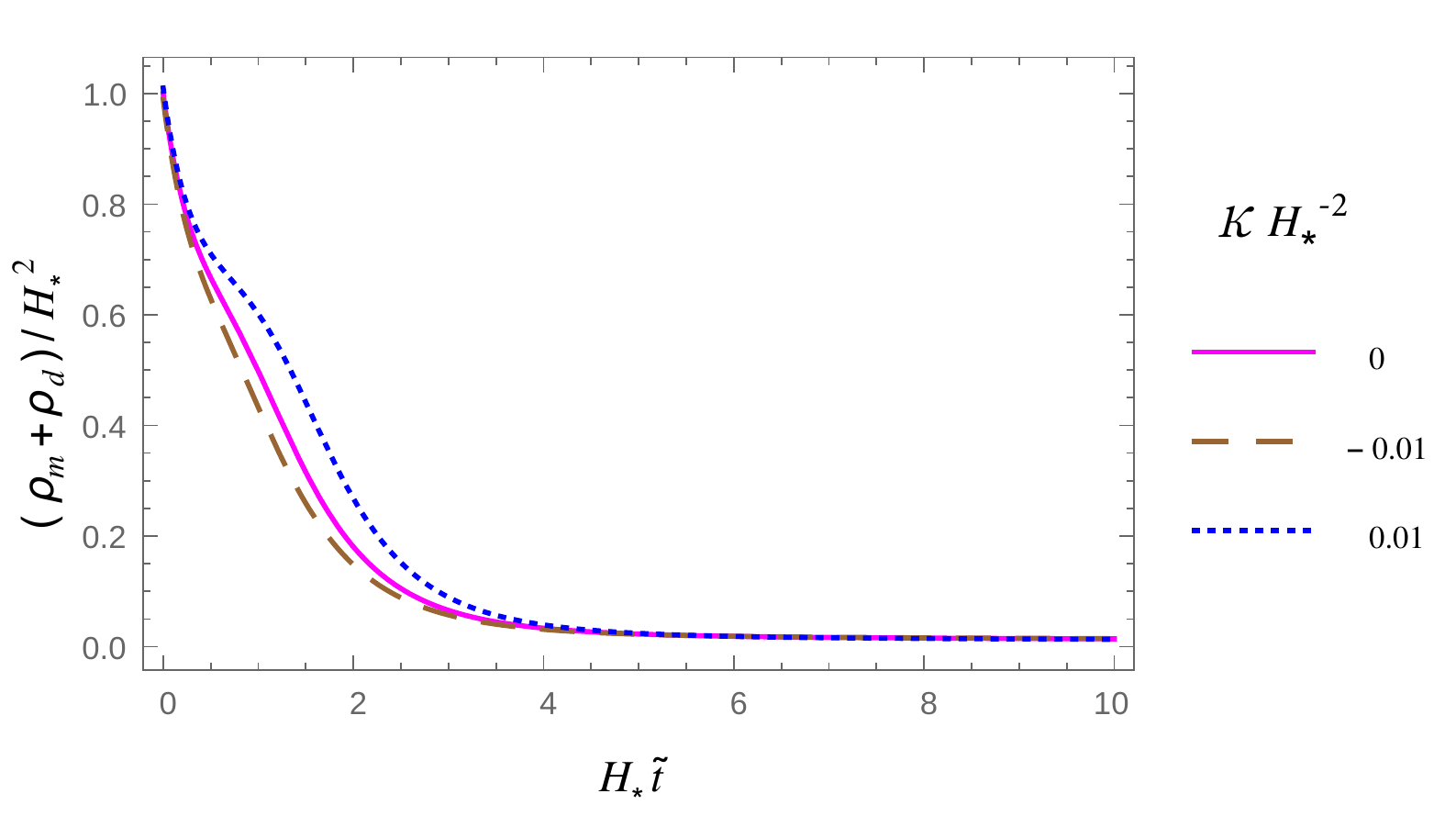}
\caption{\label{Quintrho}The variation of the total energy density with time for a Universe composed of a VdW matter distribution and Quintessence.}
\end{figure}

We thus have a system of two differential equations that can be solved numerically by again making use of the initial conditions $a(0)=\dot{a}(0)=1$, as well as $\varpi(0)=0$. The parameters are assigned values as follows: $\alpha=0.4$, $w=0.6$, $w_\text{d}=-0.9$, $\rho_\text{d}^0=0.03$ and $|\kappa|=0.01$ (or zero). The resulting plots are shown in figures $\ref{Quintacc}$, $\ref{QuintH}$, $\ref{Quintp}$ and $\ref{Quintrho}$. This time, the late-time acceleration missing in the pure VdW model shows up (Fig. $\ref{Quintacc}$), driven by the negative pressure of dark energy (Fig. $\ref{Quintp}$). Hence, adding Quintessence can be considered an improvement. 

It can be noted that the effects of curvature on the cosmic evolution during the epochs of inflation and deceleration, as well as during the graceful exit from the former to the latter, are as in the pure VdW scenario (albeit being more pronounced): with respect to a flat Universe, positive $\kappa$ delays the onset of deceleration and increases the maximum acceleration and deceleration, while negative $\kappa$ does the contrary (Fig. $\ref{Quintacc}$). A similar delay is observed in the onset of late-time acceleration: the closed Universe is the last to undergo the transition to this epoch. As stated previously, changes in the acceleration/deceleration of the Universe are caused by variations in the pressure of the cosmic fluid. In fact, the evolution of the total pressure for both the closed and open geometries is characterized by a temporal shift with respect to that for the flat Universe (Fig. $\ref{Quintp}$), mirroring the shift in the time at which the three Universes enter a new epoch. As before, small changes in the initial conditions or the parameters do not alter the effects of curvature on the cosmic history (except in magnitude), provided that the pressure of the VdW fluid still evolves smoothly. 

Another noticeable difference is the fact that the deviation from the flat Universe is stronger for positive $\kappa$. One also notes that none of the three geometries undergoes an exponential inflationary expansion (Fig. $\ref{QuintH}$). This is in contrast with the VdW-only scenario, in which the value for the Hubble parameter of the flat Universe at early times is approximately constant (see Fig. $\ref{VdWH}$), suggesting that inflation is close to being exponential.  

\begin{figure}[t]
\centering
\includegraphics[width=10cm]{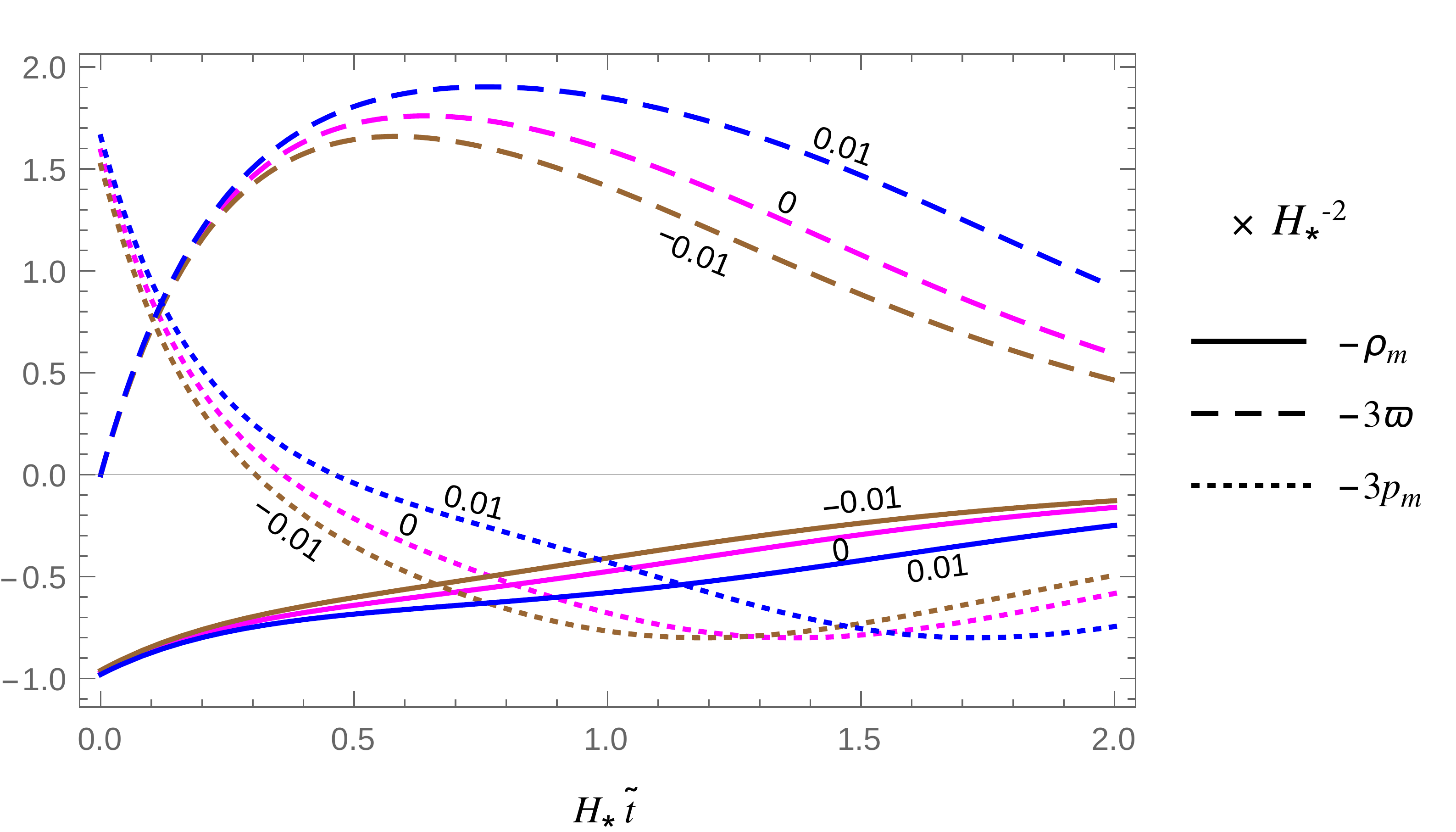}
\caption{\label{Analysis1}The three main factors contributing to the difference in the maximum inflationary acceleration of the flat, open and closed Universes. Comparison with Fig. $\ref{Quintacc}$ reveals that the primary cause of this difference is the variation of the non-equilibrium pressure $\varpi$ with $\kappa$; any change in $\rho_\text{m}$ from one geometry to another \emph{decreases} the separation between the corresponding curves of acceleration vs time. This also holds for changes in $p_\text{m}$ during the latter part of the time interval considered here. The Universe is modeled as a dissipative mixture of Quintessence and a VdW fluid, with the spatial curvature denoted by the label on each curve.}
\end{figure}

Our results are in line with the analysis presented by Kremer and Teixeira da Silva in Ref.~\refcite{Kremernew}. These authors model the Universe as a scalar field mixed with a matter field, and likewise find that the initial acceleration and subsequent deceleration are largest for the closed Universe, while the transition to the present epoch of acceleration occurs first for the open Universe. They suggest that the greater inflationary acceleration for $\kappa>0$ is due to a greater $|\varpi|$ since, being negative, the non-equilibrium pressure $\varpi$ helps to drive acceleration, as can be deduced from the equation
\begin{equation}
\ddot{a}=-\frac{a}{2}(\rho_\text{T}+3p_\text{h}+3\varpi)
\label{a''}
\end{equation}  
where $\rho_\text{T}=\rho_\text{m}+\rho_\text{d}$ and $p_\text{h}=p_\text{m}+p_\text{d}$. We have found that this is also so in our case (Fig. $\ref{Analysis1}$). Meanwhile, $\varpi$ is different in the flat, open and closed Universes mainly because of their different expansion rates. There is a strong link between $\varpi$ and $a$ -- as Eq. ($\ref{NonEqPressure}$) suggests, $|\dot{\varpi}|$ is initially largest for $\kappa>0$. Hence $\varpi$ grows more rapidly in the closed Universe, which consequently expands at the fastest rate (due to the negative pressure associated with $\varpi$). In turn, the rapid expansion causes the strong gravitational field present at inflation to decay more quickly than it does for the flat and open geometries, increasing the rate of energy transfer to the VdW fluid and thus resulting in the growth of $\varpi$. The cycle is then repeated.

\begin{figure}[b]
\centering
\includegraphics[width=10cm]{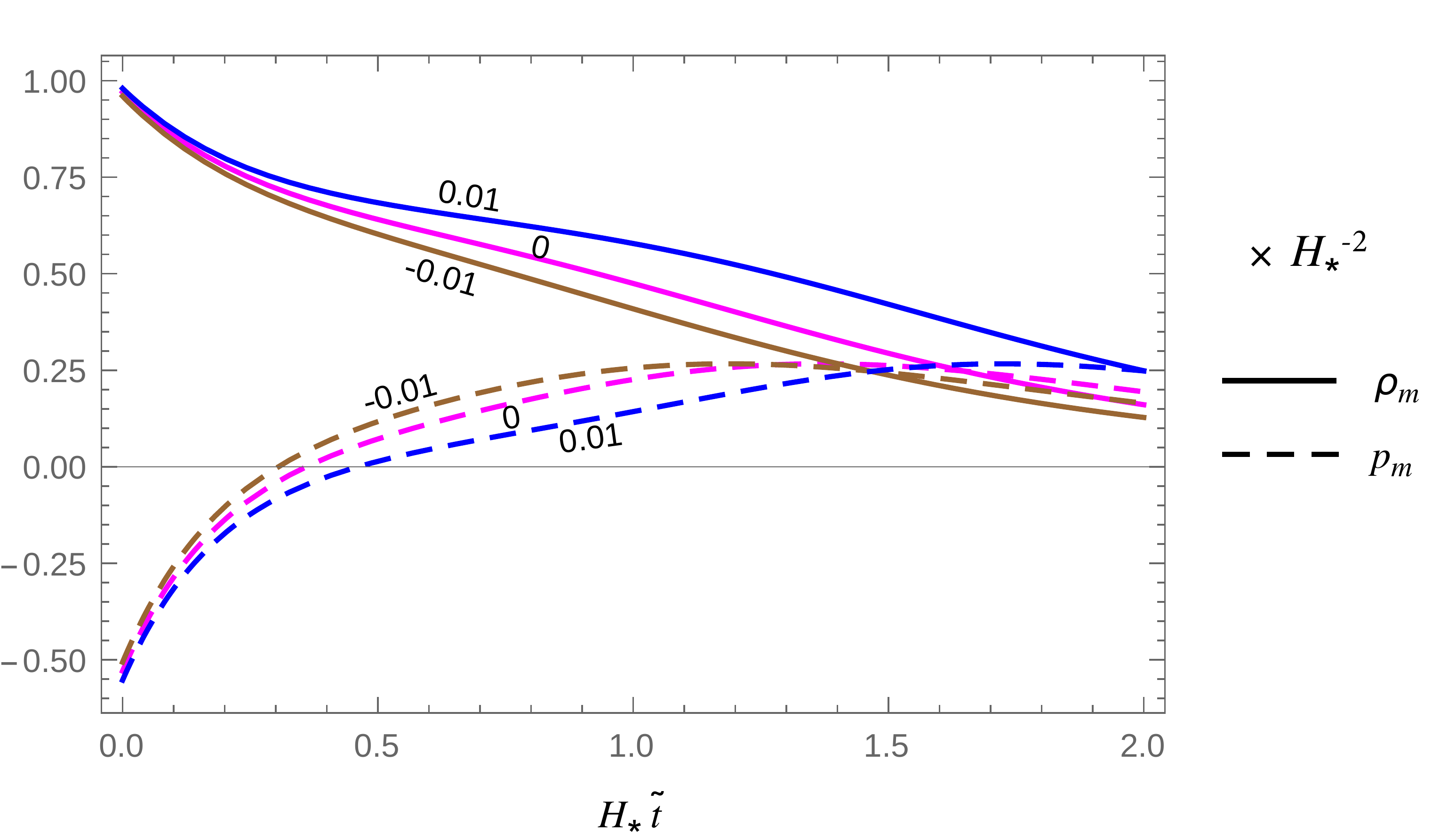}
\caption{\label{Analysis3}Curves showing how the energy density and pressure of matter vary with time. The Universe is modeled as a dissipative mixture of Quintessence and a VdW fluid. The label on each curve denotes the respective value of $\kappa$.}
\end{figure}

The fact that the repulsive pressure initially exerted by the VdW matter distribution is largest for $\kappa>0$ also helps to boost the inflationary acceleration of the closed Universe (as does the fact that $p_\text{m}$ is smallest in this Universe just after it becomes attractive). Until $\rho_\text{m}$ has decayed by a sufficient amount, the right-hand side of Eq. ($\ref{VdWEoS}$) is dominated by the term $-3\rho_\text{m}^2$. Consequently, at first the matter component behaves as an exotic fluid\footnote{We may thus still talk about a matter-dominated period of cosmic deceleration, even though the inflationary epoch in this model is also, strictly speaking, a result of the particular matter distribution.} and exerts a negative pressure, whose magnitude is largest where the fluid is most dense (hence in the closed Universe). The greater fluid density associated with positive curvature in turn stems from the significant decrease that $|\varpi|$, being largest for $\kappa>0$, effects in the rate of matter decay during the early stages of cosmic evolution (refer to Eq. ($\ref{energycons3}$)). The variation of $\rho_\text{m}$ and $p_\text{m}$ with time is depicted in Fig. $\ref{Analysis3}$.

\begin{figure}[tb]
\centering
\includegraphics[width=10cm]{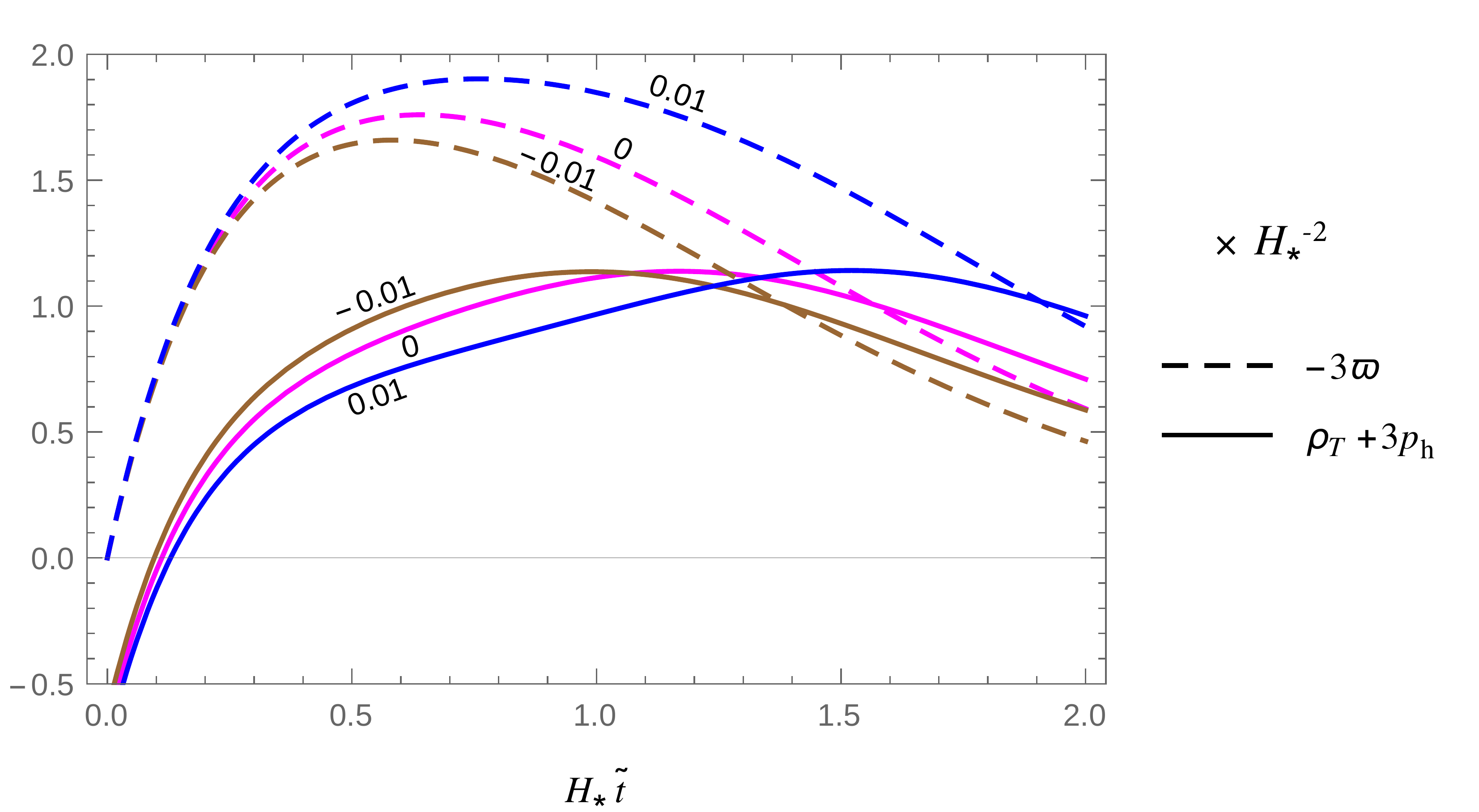}
\caption{\label{Analysis2}Curves depicting the evolution of the non-equilibrium pressure $\varpi$ in a Universe modeled as a dissipative mixture of Quintessence and a VdW fluid. Labels denote the spatial curvature $\kappa$. The second set of curves shows how the sum of the total energy density and thrice the total hydrostatic pressure varies with time. The intersection between curves with the same $\kappa$ indicates the onset of deceleration for the respective Universe.}
\end{figure}

The larger magnitude that $\varpi$ has in the closed Universe explains the delay in the transition from the first accelerated epoch to the subsequent deceleration. Eq. ($\ref{a''}$) implies that this transition occurs when $3|\varpi|$ is equal to the sum $\rho_\text{T}+3p_\text{h}$ -- in other words, when the repulsive non-equilibrium pressure ($p_\text{d}$ is negligible at this point) is counterbalanced by the attractive force of matter. The smaller $|\varpi|$ in the open Universe thus leads to an earlier onset of deceleration; the matter content would have to decay faster for the transition to the decelerated period to coincide with that of the closed Universe. This is illustrated in Fig. $\ref{Analysis2}$.
\begin{figure}[b]
\centering
\includegraphics[width=10cm]{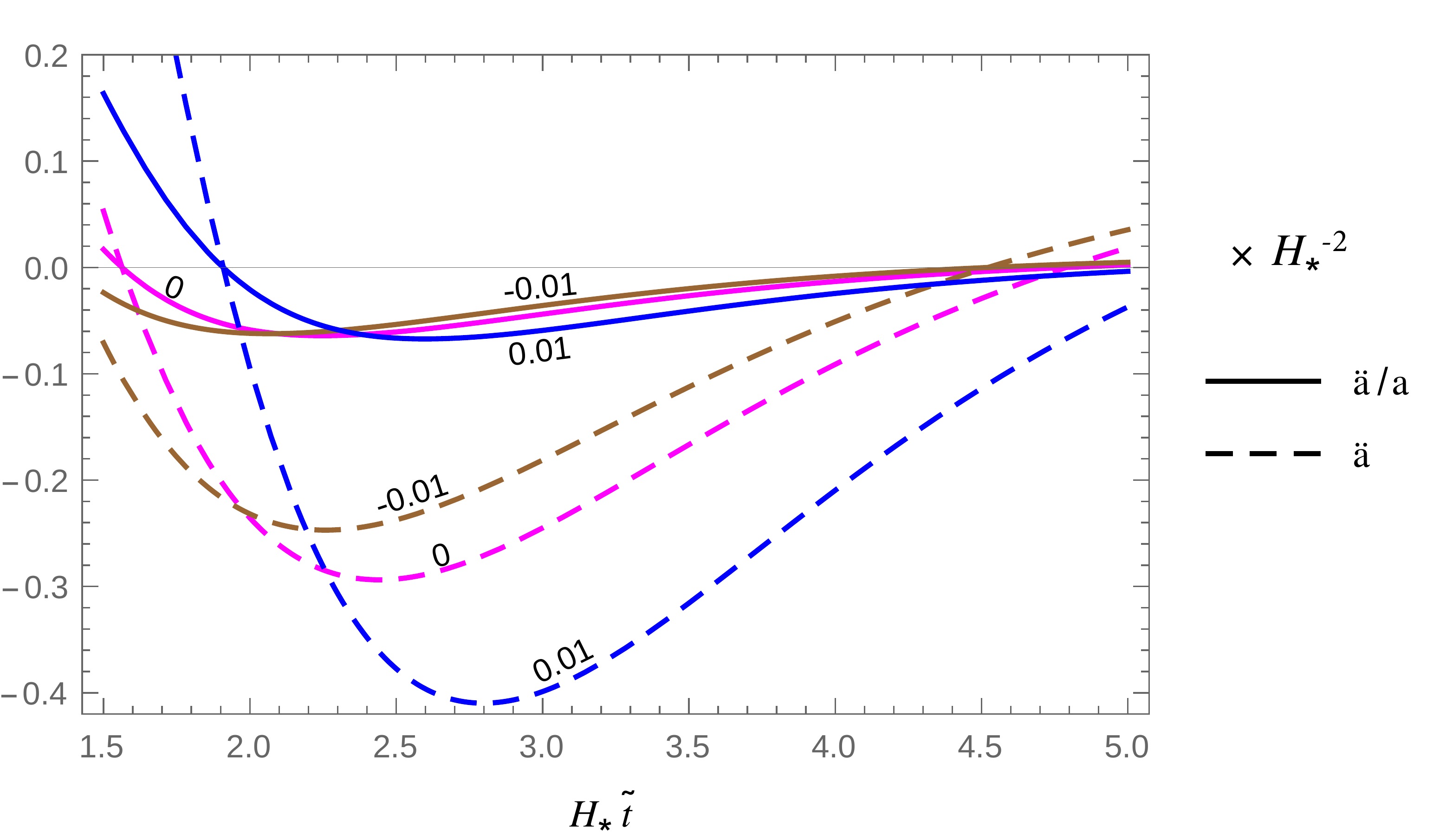}
\caption{\label{Analysis4}Curves showing the temporal variation of $\ddot{a}/a$ and $\ddot{a}$ for a Universe composed of Quintessence and a VdW fluid. Labels denote the respective value of $\kappa$.}
\end{figure}

As suggested above, Eq. ($\ref{energycons3}$) allows one to conclude that a negative non-equilibrium pressure increases $\dot{\rho}_\text{m}$ (This would constitute a decrease in $|\dot{\rho}_\text{m}|$ if $\dot{\rho}_\text{m}<0$). The authors of Ref.~\refcite{Kremernew} point out that, since $|\varpi|$ is greatest in the closed Universe, this is also where the most matter is produced, hence leading to a larger deceleration. In the case under consideration, however, things are somewhat different: although the non-equilibrium pressure decreases the rate at which matter decays, it is never large enough to yield an overall increase in its energy density. Fig. $\ref{Analysis4}$ allows us to conclude that the closed Universe undergoes the greatest deceleration mainly due to the larger value of its scale factor, since $a$ scales the sum $\rho_\text{T}+3p_\text{h}+3\varpi$ in Eq. ($\ref{a''}$). What we are essentially saying here is that, given the larger size of the closed Universe at this time, the comparability of the quantity $\rho_\text{T}+3p_\text{h}+3\varpi$ ($=-2\ddot{a}/a$) for all three values of $\kappa$ -- as evident from Fig. $\ref{Analysis4}$ -- must mean that the total amount of matter is greatest in the closed scenario (dark energy is still negligible at this point in time, and Fig. $\ref{Analysis2}$ demonstrates that $\varpi$ would have become less relevant). Consequently, the resultant repulsion is also larger, as is the deceleration it produces.

Ref.~\refcite{Kremernew} invokes the greater production of matter for $\kappa>0$ (together with the faster decay of the scalar field) to explain why the onset of late-time acceleration is delayed when the geometry is closed. We also attribute the said delay to a larger matter pressure: by the time the epoch of deceleration comes to an end, the first term on the right-hand side of $(\ref{VdWEoS})$ would have started to dominate, reducing the VdW EoS to approximately $p_\text{m}=w\rho_\text{m}$. This implies a positive pressure that is again larger where the fluid is most dense, i.e. in the closed Universe. 

The transition time to the present epoch of acceleration is also affected by the dependence of the dark energy density on $\kappa$, albeit to a much lesser extent. The negative pressure associated with dark energy gives rise to a repulsive effect and thus, as previously asserted, is the primary factor that drives late-time acceleration. Since $\rho_\text{d}$ varies in inverse proportion to the scale factor (see Eq. ($\ref{darkenergydensity}$)), it is smallest in the closed Universe, for which $a$ is largest. Consequently, this is also where dark energy exerts the least pressure at any given time, contributing to the delay in the onset of late-time acceleration.

\subsubsection{Chaplygin gas}
We now model dark energy as a Chaplygin gas, whose EoS reads:\cite{Kremer}
\begin{equation}
p_\text{d} = -\frac{A}{\rho_\text{d}};~~~A>0
\label{ChaplyginEoS}
\end{equation}
where $A$ is a constant. This can be inserted into Eq. ($\ref{energycons2}$), which is then solved for $\rho_\text{d}$, yielding:
\begin{equation}
\rho_\text{d}=\frac{\rho_\text{d}^0}{a^3}~\sqrt{\frac{a^6+\psi}{1+\psi}}
\label{darkenergydensity2}
\end{equation}
with $\rho_\text{d}^0$ representing the dark energy density at $\tilde{t} = 0$; $\psi$ is a dimensionless quantity inversely proportional to $A$ and related to $\rho_\text{d}^0$ and $A$ via the equation 
\begin{equation}
\psi = \frac{1}{A}(\rho_\text{d}^0)^2-1.
\label{A}
\end{equation}
\begin{figure}[t]
\centering
\includegraphics[width=10cm]{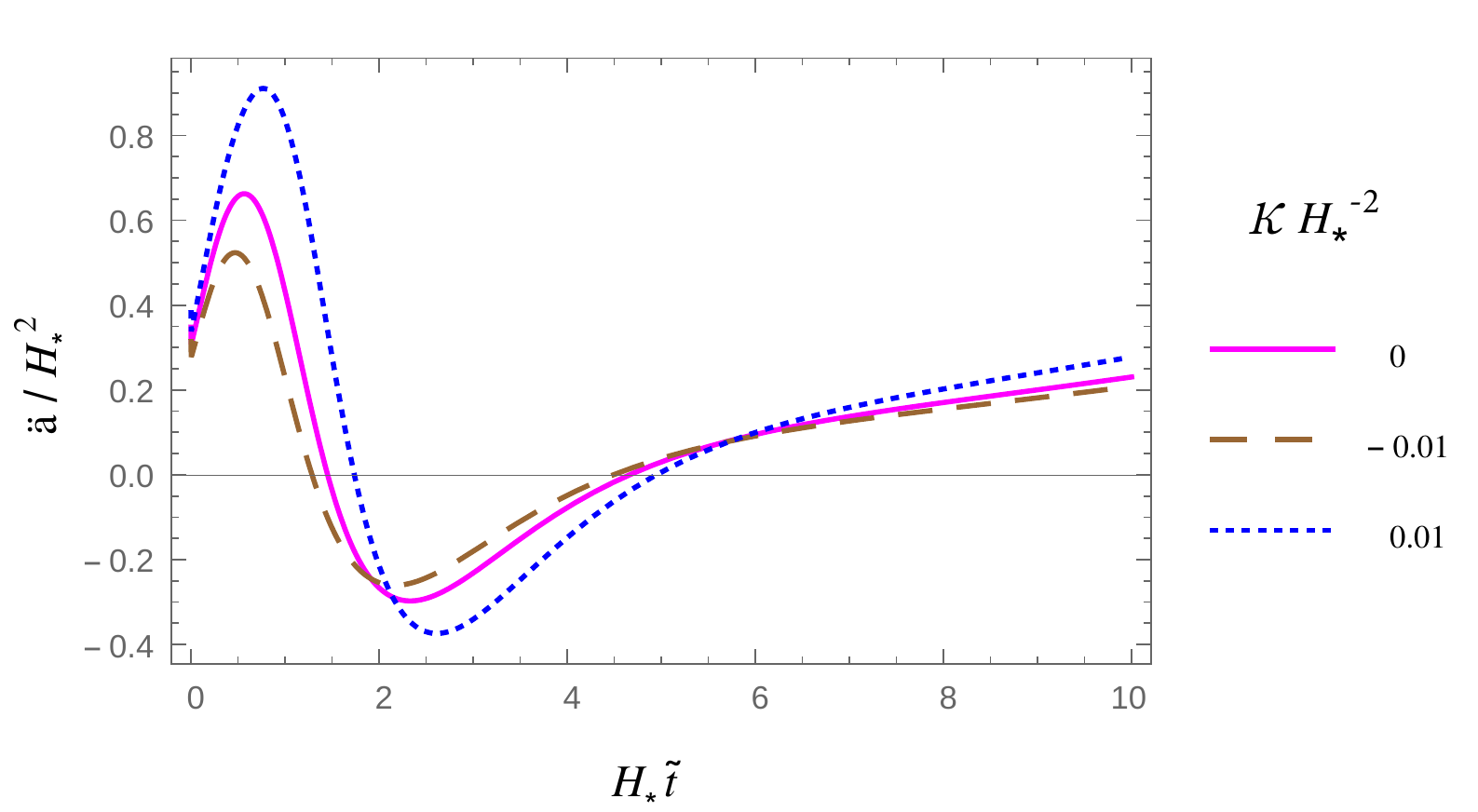}
\caption{\label{Chapacc}The variation of acceleration with time for a Universe composed of a VdW matter distribution and dark energy with a Chaplygin gas EoS.}
\end{figure}

We repeat the procedure adopted in the case of Quintessence and take the time derivative of Eq. ($\ref{1stFriedmann2}$), then substitute for $\dot{\rho}_\text{d}$ and $\dot{\rho}_\text{m}$ from equations ($\ref{energycons2}$) and ($\ref{energycons3}$), respectively. The pressure of matter is determined by the VdW EoS (Eq. ($\ref{VdWEoS}$)), and that of dark energy by ($\ref{ChaplyginEoS}$), with $A$ written in terms of $\rho_\text{d}^0$ and $\psi$ via Eq. ($\ref{A}$). The matter energy density $\rho_\text{m}$ is subsequently replaced with the corresponding expression from Eq. ($\ref{1stFriedmann2}$), while $\rho_\text{d}$ takes the form specified in ($\ref{darkenergydensity2}$). The final equation reads:
\begin{align}
&3h(\tilde{t},\kappa)\left[\frac{8wa^3}{a(3a^2-\kappa-\dot{a}^2)+g(\tilde{t}) \rho_\text{d}^0}-\frac{3h(\tilde{t},\kappa)}{a^2}\right]+3a^2\varpi-\frac{3\rho_\text{d}^0 a^5}{g(\tilde{t}) (1+\psi)}+\kappa+\dot{a}^2\notag\\+~&2a\ddot{a}=0
\label{MainEq3}
\end{align}
where
\begin{equation}
g(\tilde{t}) = \sqrt{\frac{a^6+\psi}{1+\psi}}~~~~~~~~~~h(\tilde{t},\kappa) = -\rho_\text{d}^0~\frac{g(\tilde{t})}{a}+\dot{a}^2+\kappa.
\end{equation}
Eq. ($\ref{MainEq3}$) also reduces to the corresponding one (Eq. (16)) in Ref.~\refcite{Kremer} when $\kappa$ is set equal to zero and the necessary modifications in notation are made. Together with ($\ref{NonEqPressure}$), Eq. ($\ref{MainEq3}$) is solved by setting $a(0)=\dot{a}(0)=1$, $\varpi(0) = 0$, $\alpha = 0.4$, $w = 0.6$, $\psi = 3$, $\rho_\text{d}^0 = 0.03$ and $|\kappa|=0.01$ (or zero if the Universe is flat).

\begin{figure}[t]
\centering
\includegraphics[width=10cm]{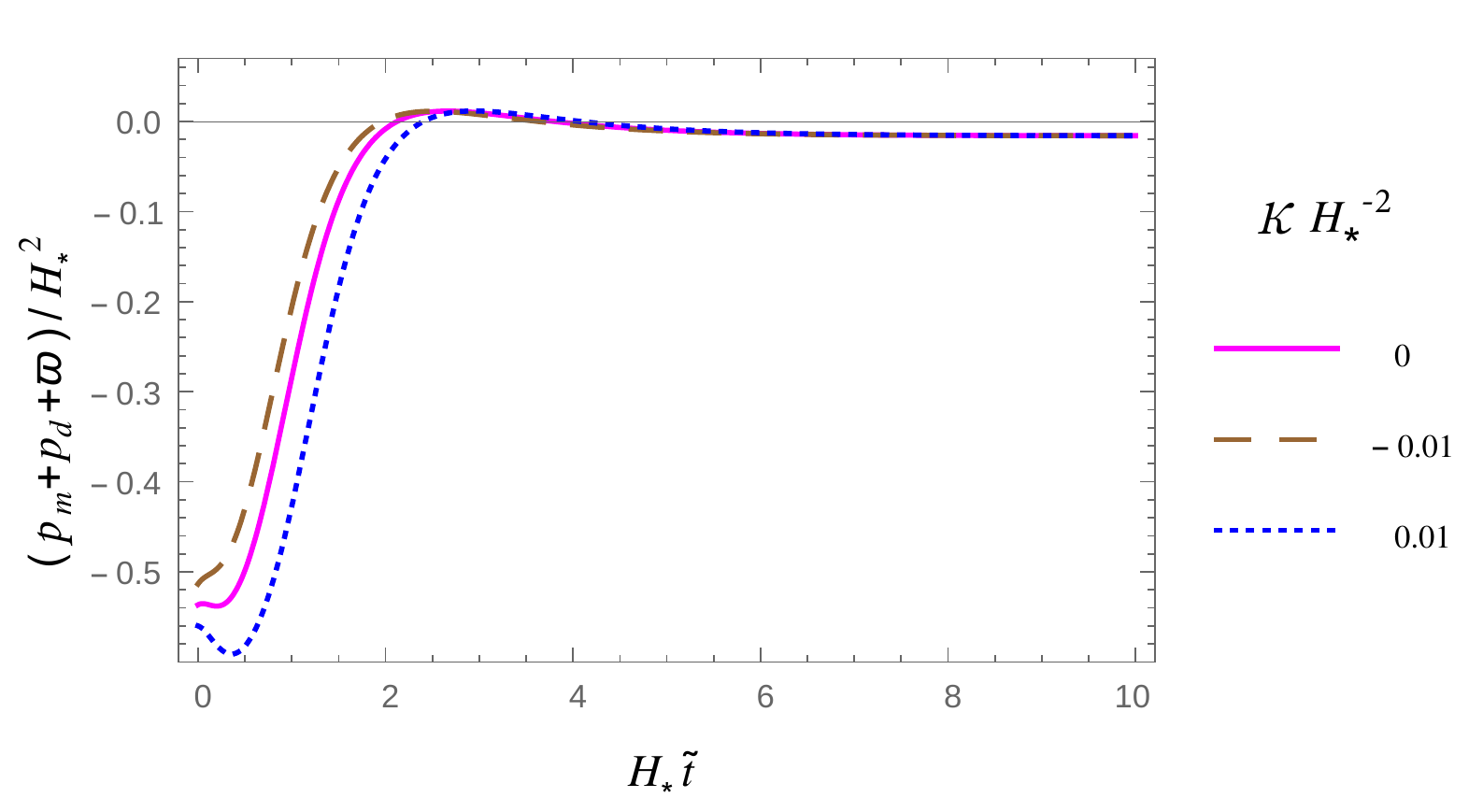}
\begin{picture}(0,0)
\put(-172.5,38){\includegraphics[height=2.5cm]{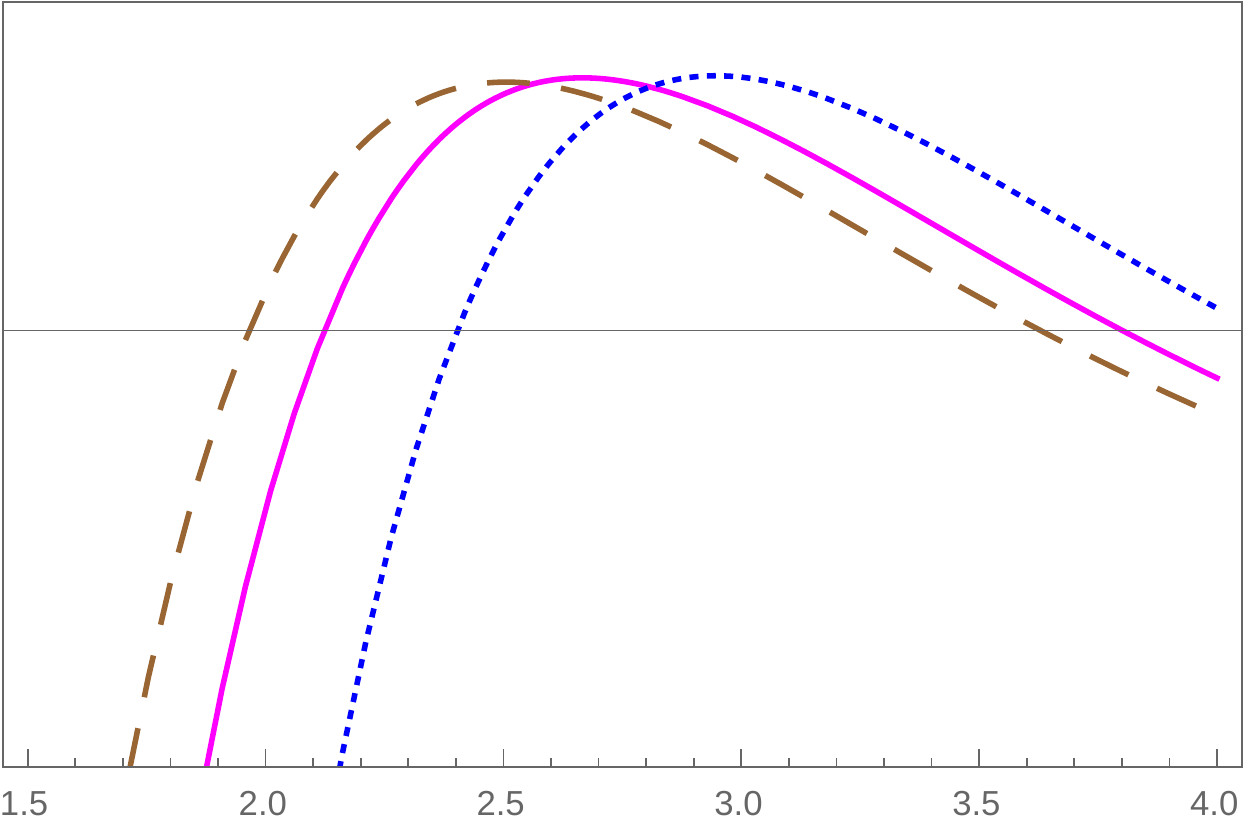}}
\end{picture}
\caption{\label{Chapp}The variation of the total pressure with time for a Universe composed of a VdW matter distribution and dark energy with a Chaplygin gas EoS. The peaks of the curves are shown at greater resolution (inset).}
\end{figure}

\begin{figure}
\centering
\includegraphics[width=10cm]{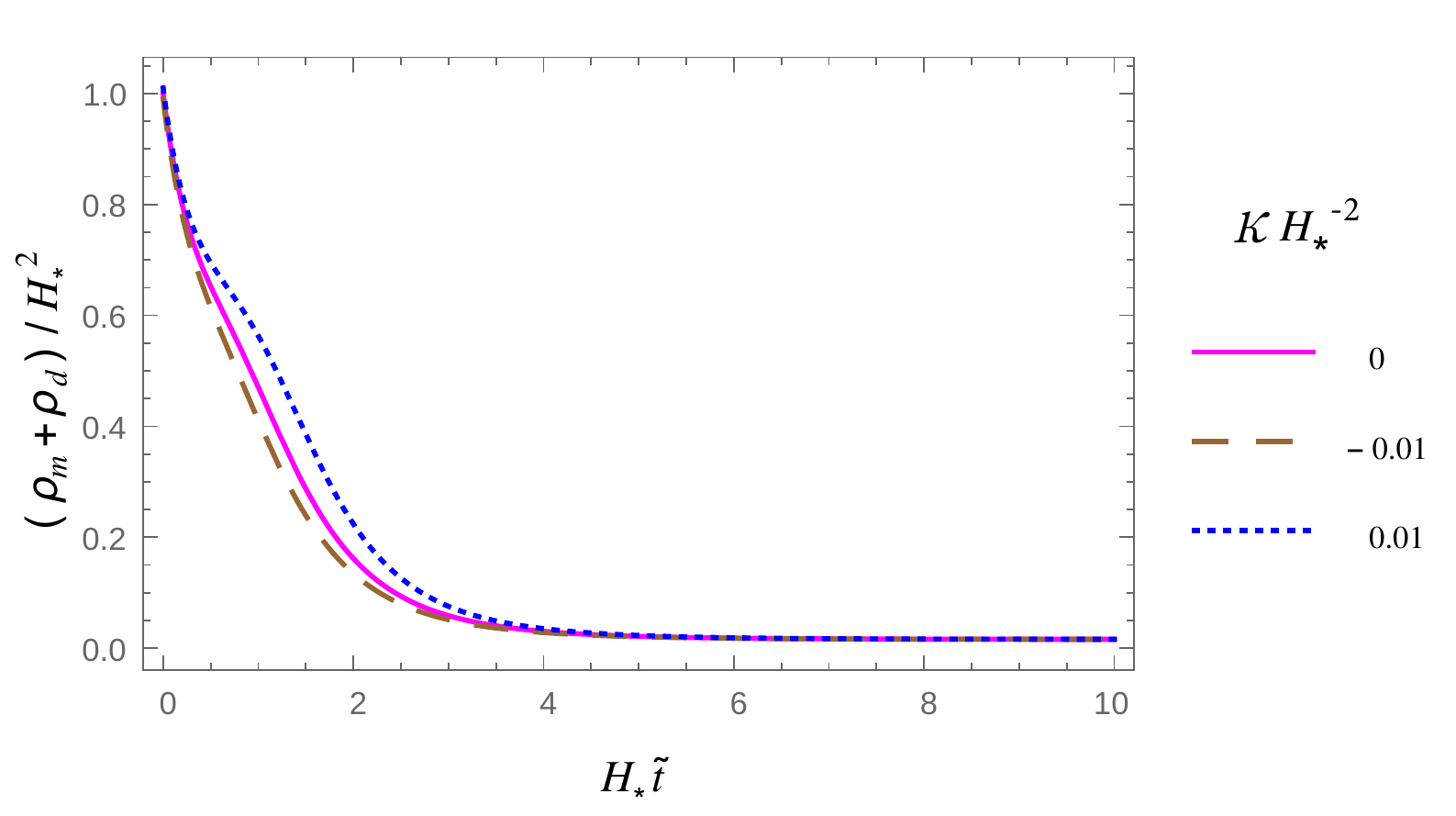}
\caption{\label{Chaprho}The variation of the total energy density with time for a Universe composed of a VdW matter distribution and dark energy with a Chaplygin gas EoS.}
\end{figure}

The evolution of the Universe with matter modeled as a VdW fluid and dark energy as a Chaplygin gas is illustrated in figures $\ref{Chapacc}$, $\ref{Chapp}$ and $\ref{Chaprho}$. It can be seen that the effects of curvature on the cosmic history are similar to those noted for the previous model, in which dark energy was modeled as Quintessence. Once again, small variations in the parameters or the initial conditions only change the magnitude of these effects.

\subsection{Dark energy as a dynamical $\boldmath{\Lambda}$}
In this section, we again consider hydrostatic pressures only, and model dark energy as a time-dependent cosmological term $\Lambda(\tilde{t})$ with EoS parameter $w_\Lambda=p_\Lambda/\rho_\Lambda=-1$ ($p_\Lambda$ and $\rho_\Lambda$ being, respectively, the pressure and energy density associated with $\Lambda(\tilde{t})$). Thus, the Friedmann equation introduced in ($\ref{1stFriedmann}$) now reads:\footnote{Note that $\nicefrac{1}{3}\Lambda(\tilde{t}) = \rho_\Lambda$.}
\begin{equation}
H^2=\rho_\text{m}-\frac{\kappa}{a^2}+\frac{\Lambda(\tilde{t})}{3}
\label{1stFriedmann3}
\end{equation}
while energy conservation can be expressed as:
\begin{equation}
\dot{\rho}_\text{m}=-3H(\rho_\text{m}+p_\text{m})-\frac{\dot{\Lambda}}{3}.
\label{energycons4}
\end{equation}
In this case, the matter and dark energy components cannot be conserved separately, because Einstein's field equations take the form\footnote{$R_{\mu\nu}$ stands for the Ricci tensor (defined in terms of the Riemann tensor as $R^\sigma_{\phantom{\mu}\mu\sigma\nu}$), $\mathcal{R}$ the Ricci scalar and $T_{\mu\nu}^\text{m}$ the matter part of the energy--momentum tensor. We use the metric signature $(-,+,+,+)$.}
\begin{equation}
G_{\mu\nu}=R_{\mu\nu}-\nicefrac{1}{2}\mathcal{R}g_{\mu\nu}=3T_{\mu\nu}^\text{m}-\Lambda(\tilde{t}) g_{\mu\nu}
\end{equation}
and hence the Bianchi identity, $\nabla^\mu G_{\mu\nu}=0$, implies that $3\nabla^\mu T_{\mu\nu}^\text{m}-g_{\mu\nu}\nabla^\mu \Lambda(\tilde{t}) = 0$. Consequently, if matter were conserved independently of $\Lambda(\tilde{t})$, the latter would have to be a constant, which is not what we want here. 

Taking the time derivative of ($\ref{1stFriedmann3}$) and using Eq. ($\ref{energycons4}$) to substitute for $\dot{\rho}_\text{m}+\nicefrac{1}{3}\dot{\Lambda}$ yields:
\begin{equation}
3a^2(p_\text{m}+\rho_\text{m})-2(\kappa+\dot{a}^2)+2a\ddot{a}=0.
\label{timederivative}
\end{equation}
We now proceed by choosing a proper model for the matter distribution and an expression for $\Lambda(\tilde{t})$.

\subsubsection{$\Lambda$ proportional to $\boldmath{a^{-m}}$}
\begin{figure}[b]
\centering
\includegraphics[width=10cm]{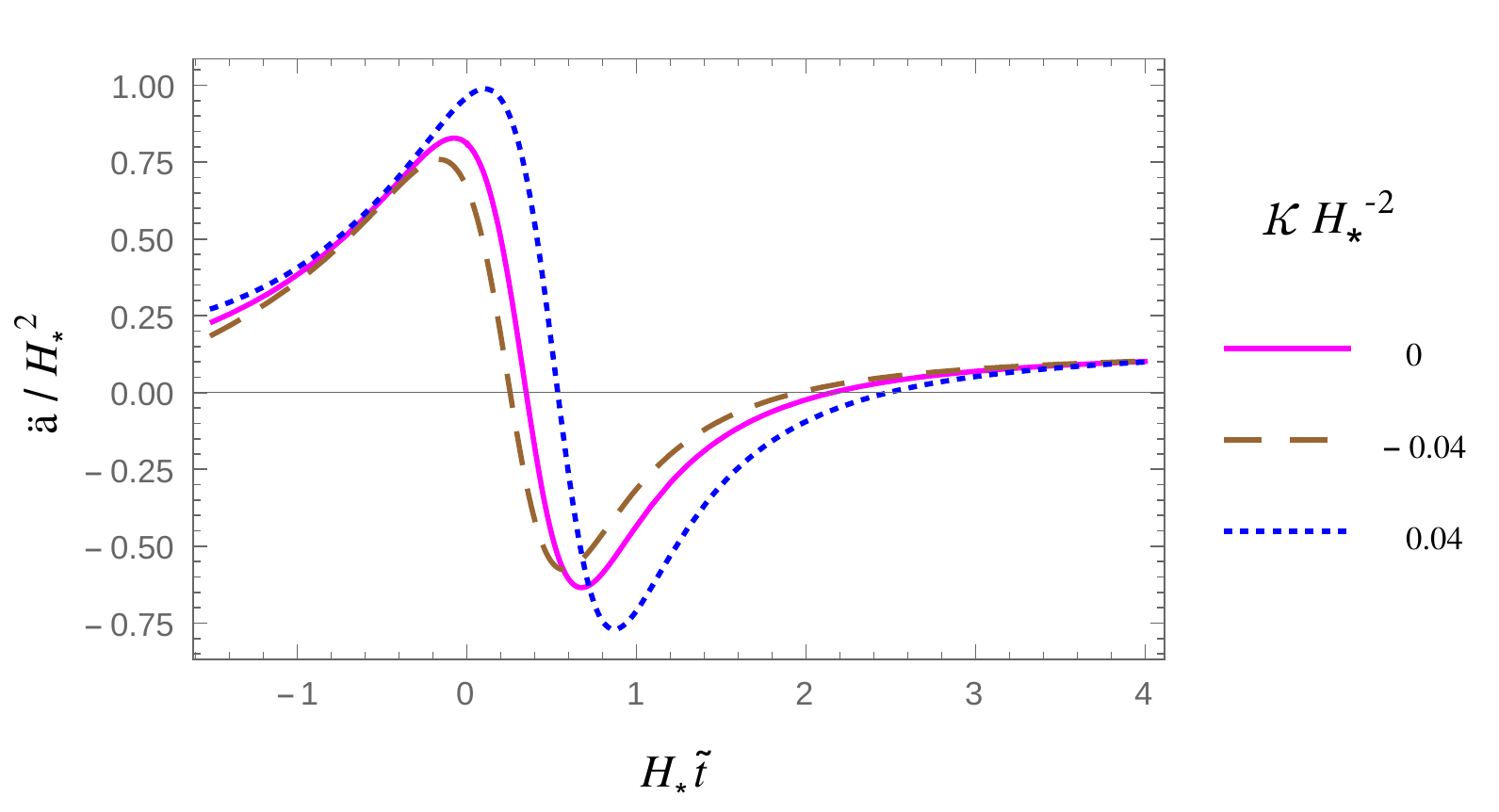}
\caption{The variation of acceleration with time for a Universe composed of a VdW matter distribution and a dynamical $\Lambda \propto a^{-1/2}$.}
\label{Lambda2acc}
\end{figure}
In the first model with a dynamical $\Lambda$, we retain the VdW EoS (Eq. ($\ref{VdWEoS}$)) for the matter part of the mixture making up the Universe, and represent the time-dependent vacuum term by:\cite{Jafarizadeh}
\begin{equation}
\Lambda(\tilde{t})=\frac{\Lambda_{\text{I}}}{a^m}
\label{Lambda1}
\end{equation}
where the parameter $m$ is allowed to take values in the range $[0,2]$ and $\Lambda_{\text{I}}$ corresponds to the value of $\Lambda(\tilde{t})$ at $\tilde{t}=0$.
\begin{figure}
\centering
\includegraphics[width=10cm]{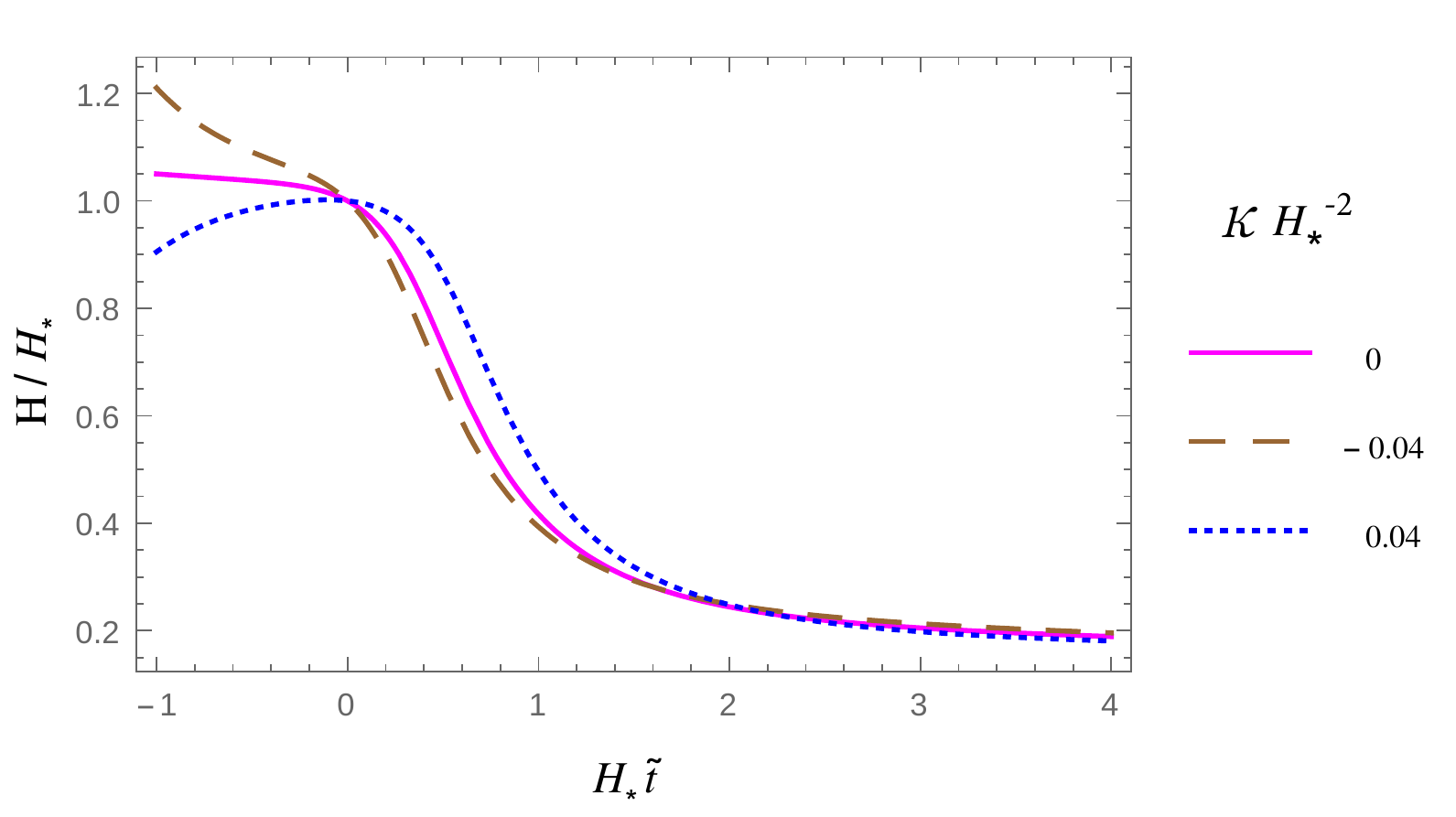}
\caption{\label{Lambda2H}The variation of the Hubble parameter with time for a Universe composed of a VdW matter distribution and a dynamical $\Lambda \propto a^{-1/2}$.}
\end{figure}
\begin{figure}[]
\centering
\includegraphics[width=10cm]{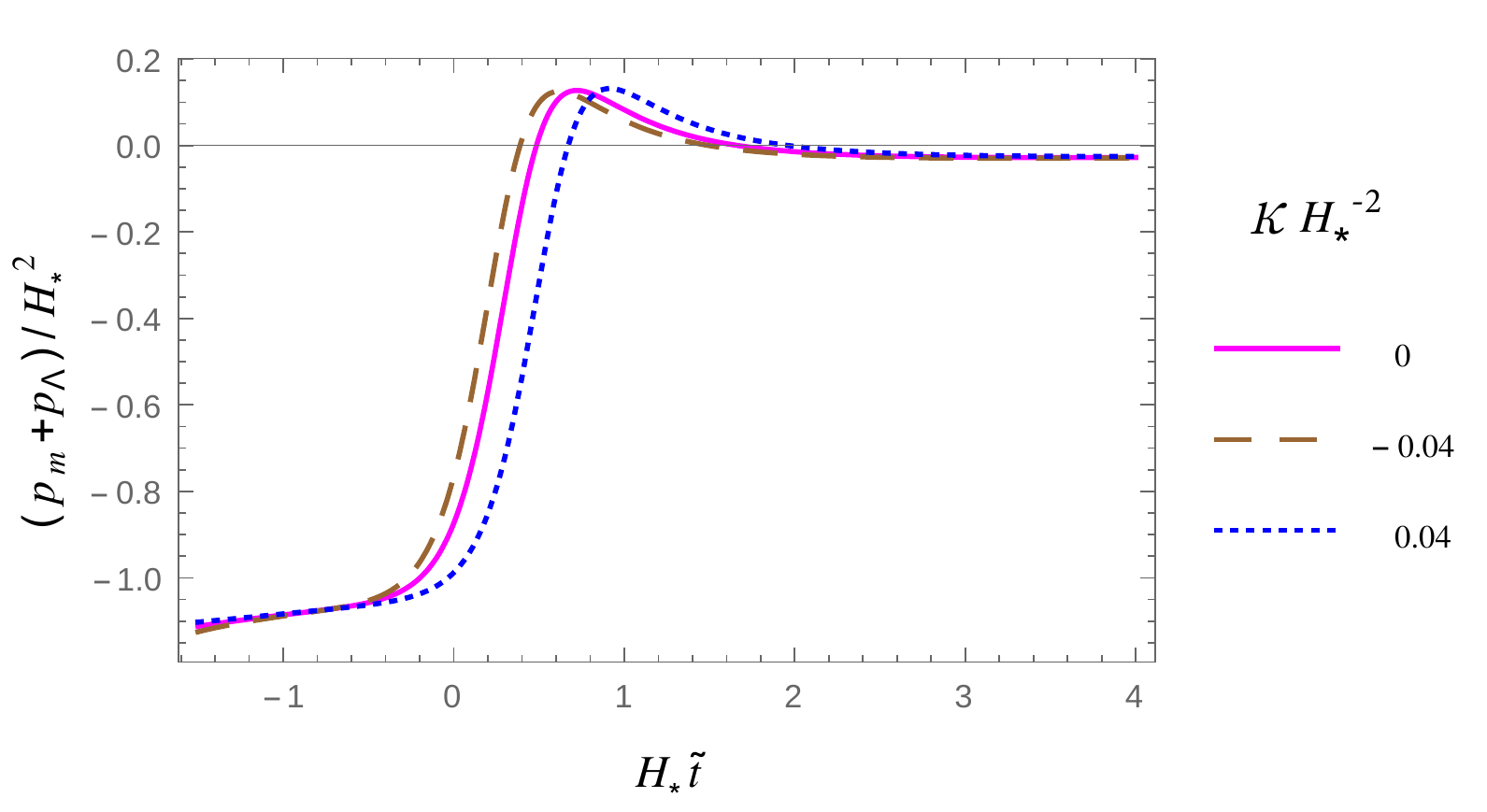}
%\begin{picture}(0,0)
%\put(-172.5,38){\includegraphics[height=2.5cm]{pics/Lambda2pzoom.pdf}}
%\end{picture}
\caption{\label{Lambda2p}The variation of the total pressure with time for a Universe composed of a VdW matter distribution and a dynamical $\Lambda \propto a^{-1/2}$.}
\end{figure} 

\begin{figure}[t]
\centering
\includegraphics[width=10cm]{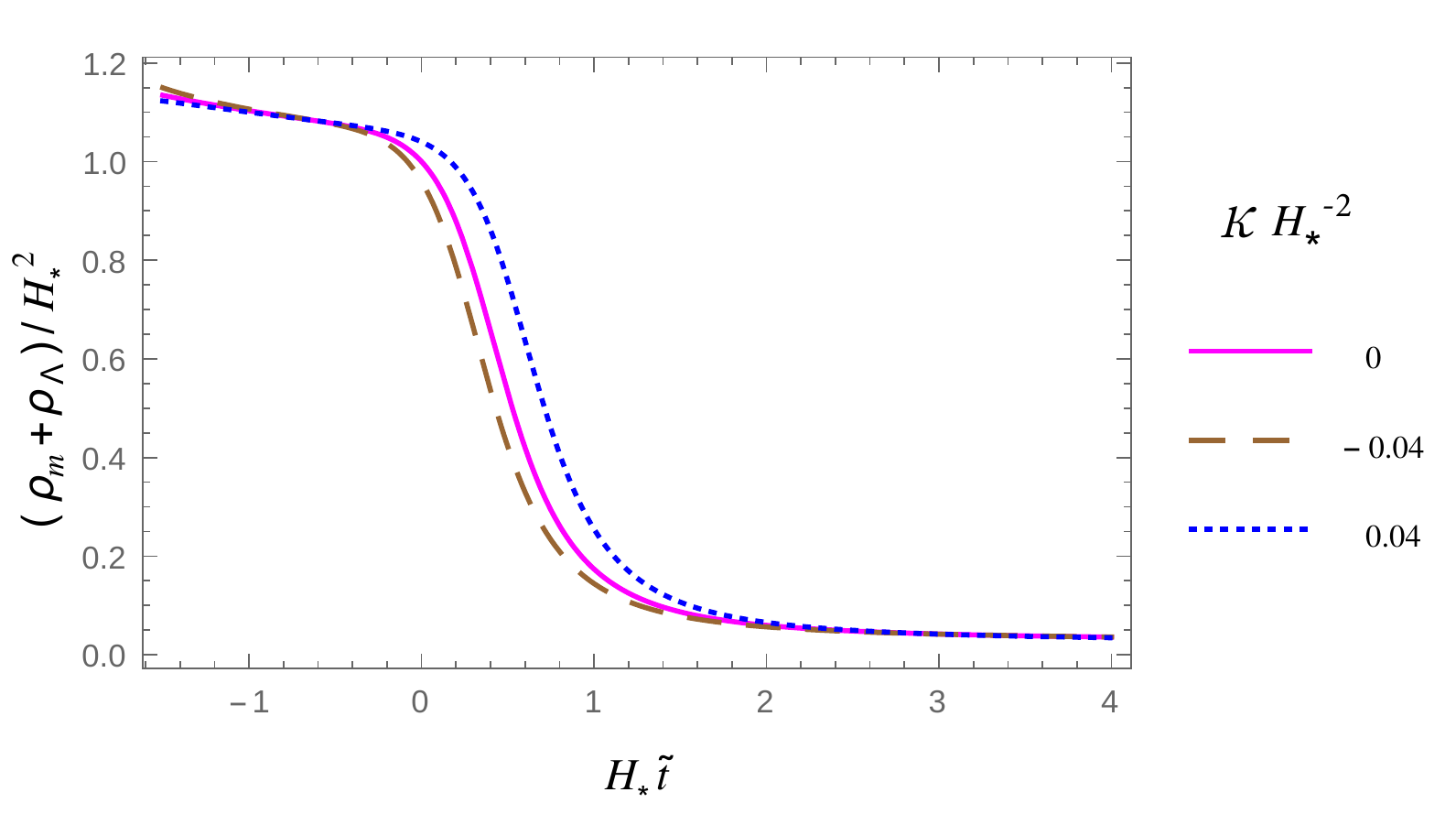}
\caption{\label{Lambda2rho}The variation of the total energy density with time for a Universe composed of a VdW matter distribution and a dynamical $\Lambda \propto a^{-1/2}$.}
\end{figure}

The fundamental equation is obtained by making use of ($\ref{VdWEoS}$) to write $p_\text{m}$ in terms of $\rho_\text{m}$, then replacing the latter with the corresponding expression from ($\ref{1stFriedmann3}$). When ($\ref{Lambda1}$) is inserted into the resulting equation, we get:  
\begin{equation}
\kappa+\dot{a}^2-\frac{\Lambda_\text{I}}{a^{m-2}}-\left[\frac{s(\tilde{t},\kappa)}{a}\right]^2-\frac{24wa^2s(\tilde{t},\kappa)}{9a^2+s(\tilde{t},\kappa)}+2a\ddot{a}=0;
\label{MainEq4}
\end{equation}
\begin{equation}
s(\tilde{t},\kappa)=\frac{\Lambda_\text{I}}{a^{m-2}}-3(\kappa+\dot{a}^2).
\end{equation}
With the initial conditions $a(0)=\dot{a}(0)=1$ and the parameter values $w=0.5$, $m=0.5$, $\Lambda_\text{I}=\num{0.2}$ and $|\kappa|=0.04$ (or zero), Eq. ($\ref{MainEq4}$) yields the cosmic history summarized in figures $\ref{Lambda2acc}$, $\ref{Lambda2H}$, $\ref{Lambda2p}$ and $\ref{Lambda2rho}$. It can be seen that the effects of curvature tally with what was observed for the previous models. Once again, these effects only change in magnitude under slight variations in the initial conditions or the parameters. This is subject to the condition that the evolution of the pressure exerted by the VdW fluid remains smooth.

\begin{figure}[b]
\centering
\includegraphics[width=10cm]{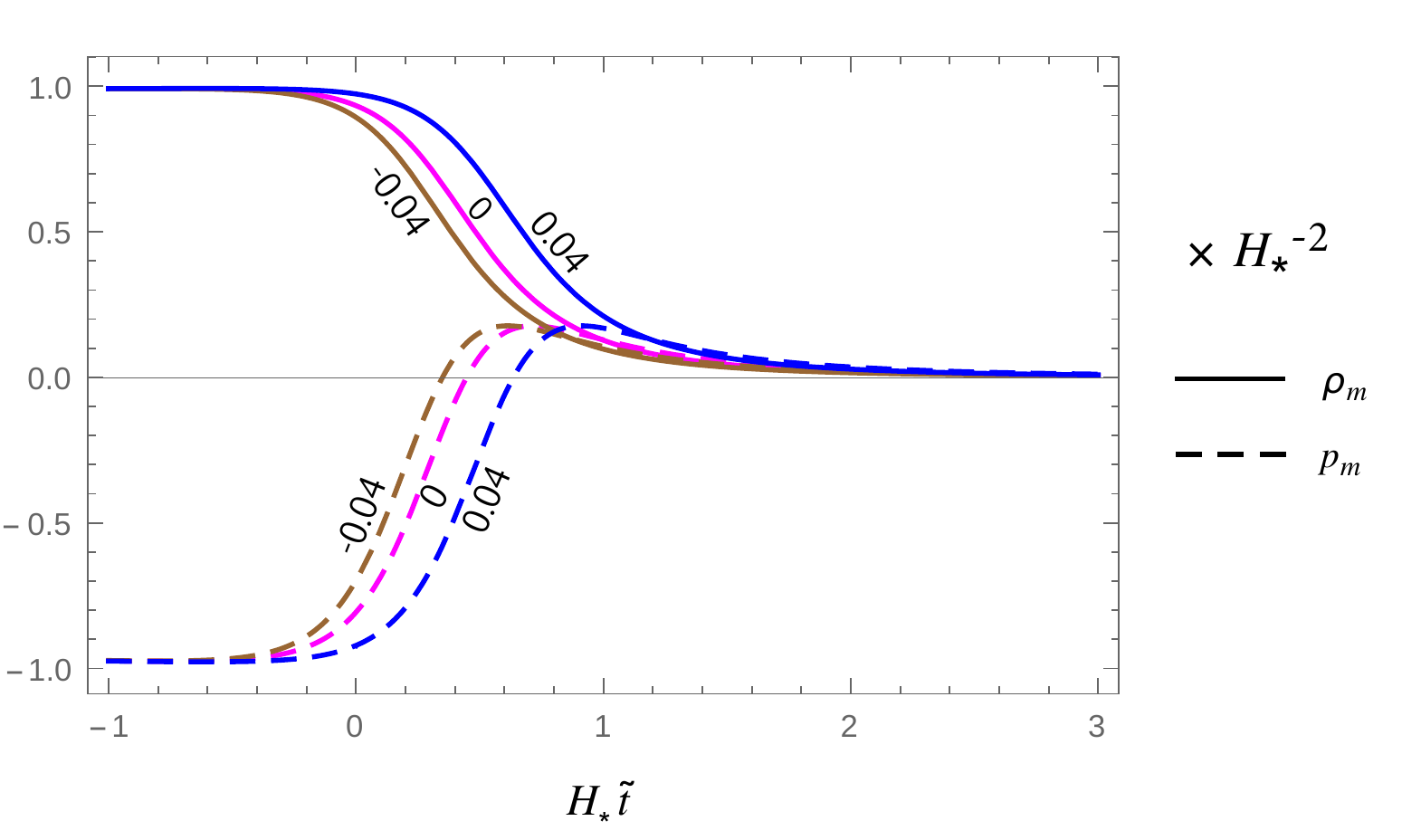}
\caption{\label{Analysis5}The temporal variation of the energy density and pressure of matter in a Universe comprising a VdW fluid and a dynamical $\Lambda \propto a^{-1/2}$. The label on each curve denotes the respective value of $\kappa$.}
\end{figure} 

We now carry out an analysis similar to the one for the VdW--Quintessence model, and try to figure out why it is that the deceleration and inflationary acceleration are largest for the closed Universe. The fact that positive curvature delays the onset of deceleration and late-time acceleration also requires clarification. 

\begin{figure}[b]
\centering
\includegraphics[width=10cm]{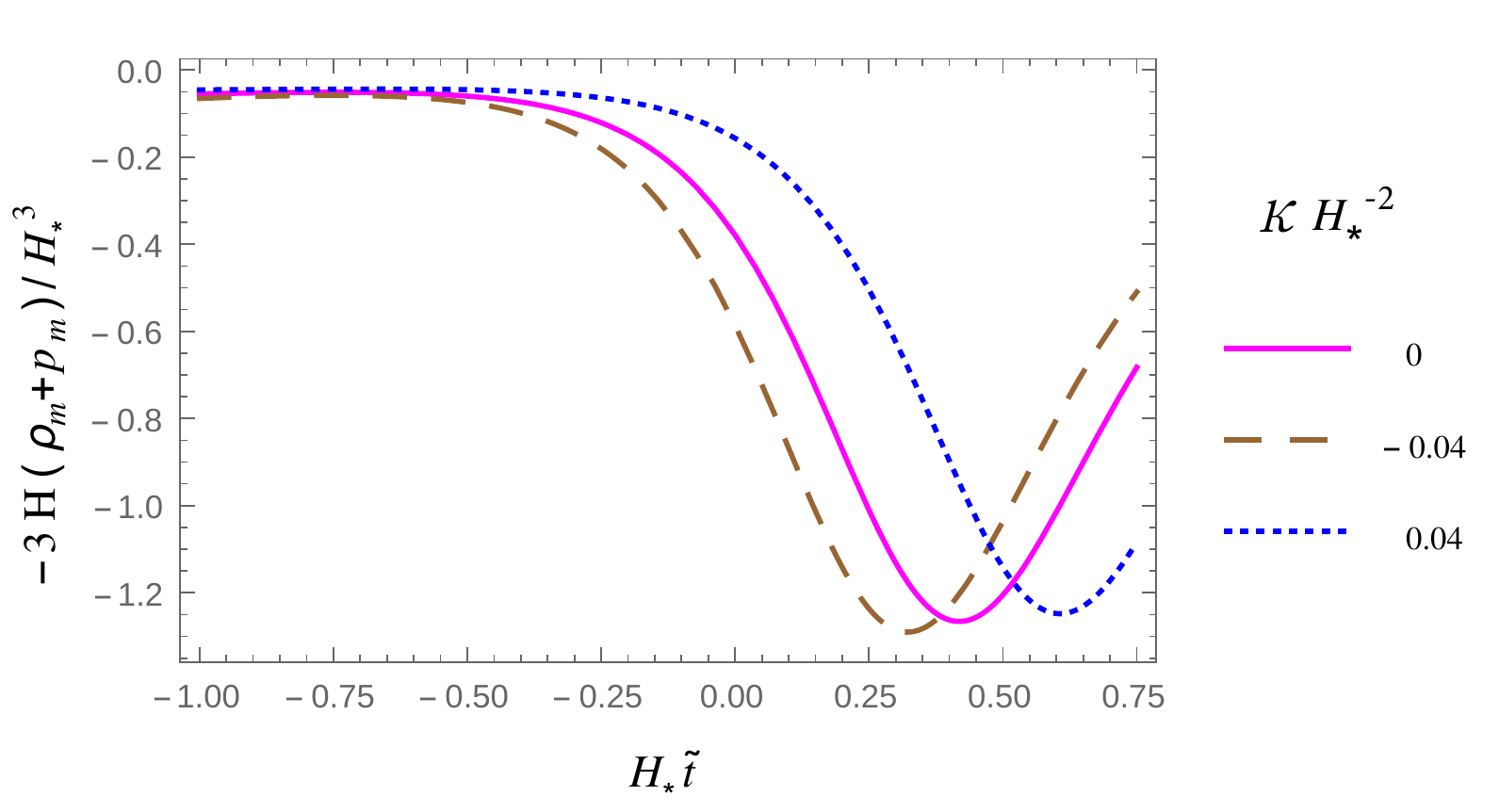}
\caption{\label{Analysis6}The evolution of $-3H(\rho_\text{m}+p_\text{m})$ during inflation and (approximately) the first half of the matter-dominated epoch. This quantity is effectively equal to $\dot{\rho}_\text{m}$ over the time domain in question. The Universe is modeled as a mixture of a VdW fluid and a dynamical $\Lambda \propto a^{-1/2}$.}
\end{figure}
\begin{figure}[b]
\centering
\includegraphics[width=10cm]{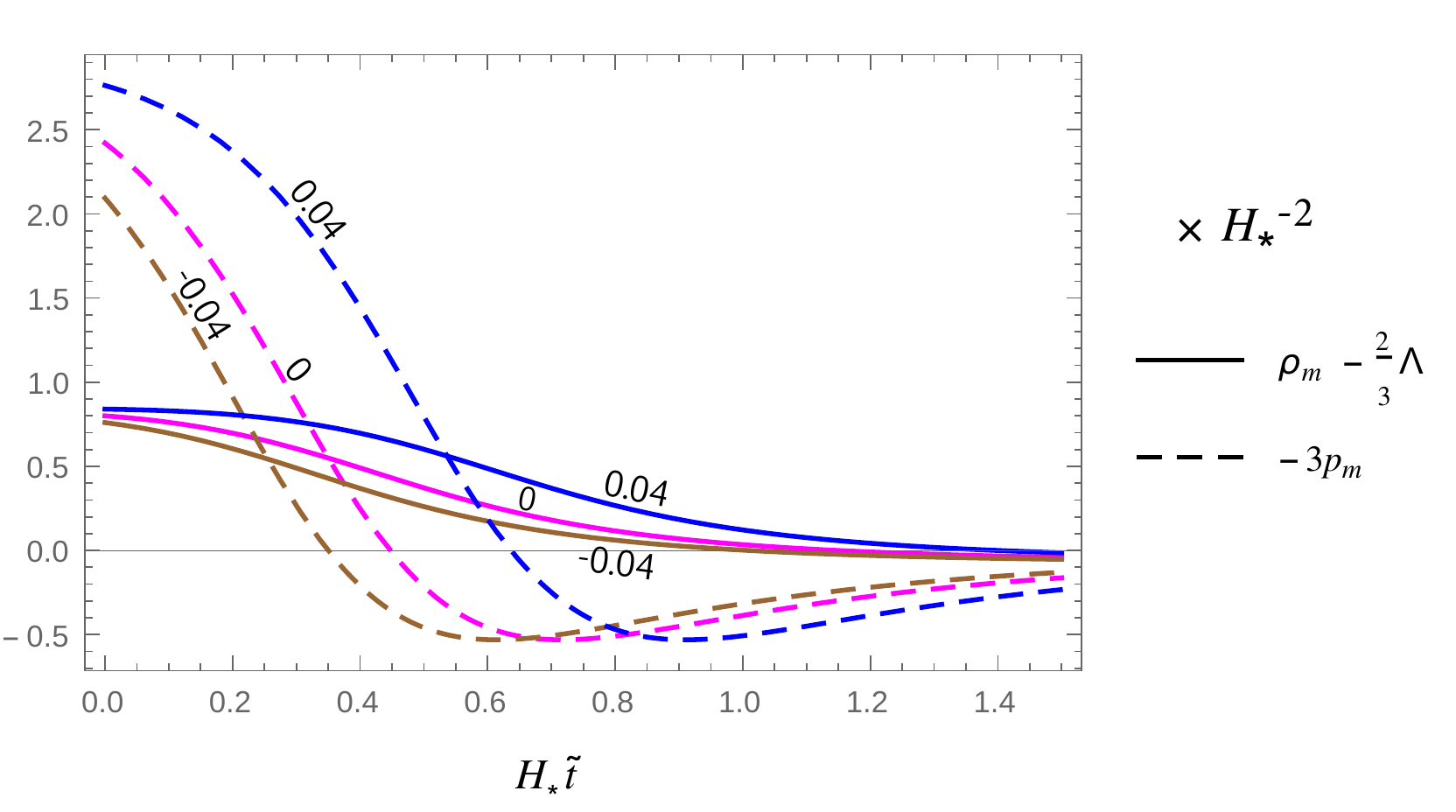}
\caption{\label{Analysis7}The evolution of $\rho_\text{m}-\nicefrac{2}{3}\Lambda$ and $-3p_\text{m}$ for a Universe composed of a VdW fluid and a dynamical $\Lambda \propto a^{-1/2}$. According to Eq. ($\ref{a''2}$), when these two quantities are equal, $\ddot{a}=0$. Hence, the point of intersection between curves with the same value of $\kappa$ (as indicated by the label on each curve) denotes the onset of deceleration for the respective Universe.}
\end{figure} 

Initially, the second term on the right-hand side of ($\ref{VdWEoS}$) dominates the dynamics of the VdW matter distribution, causing it to behave as an exotic fluid and exert a negative pressure. Eq. ($\ref{1stFriedmann3}$) implies that at $\tilde{t}=0$, $\rho_\text{m}$ is greatest in the closed Universe, for which $\kappa>0$. Consequently, this is also where the magnitude of the pressure associated with the VdW fluid is largest (Fig. $\ref{Analysis5}$), and hence where the resulting repulsive effect is most significant. In fact, at $\tilde{t}=0$ the closed Universe undergoes the biggest acceleration. Thus it soon starts to expand at the fastest rate. 

According to Eq. ($\ref{energycons4}$), the larger repulsive pressure associated with positive curvature also means a slower rate of matter decay (Fig. $\ref{Analysis6}$), so that the VdW fluid remains densest in the closed Universe. This again implies a greater repulsion (and acceleration), providing an explanation for the different amplitudes of the acceleration--time curves (Fig. $\ref{Lambda2acc}$) at the end of inflation.

\begin{figure}[b]
\centering
\includegraphics[width=10cm]{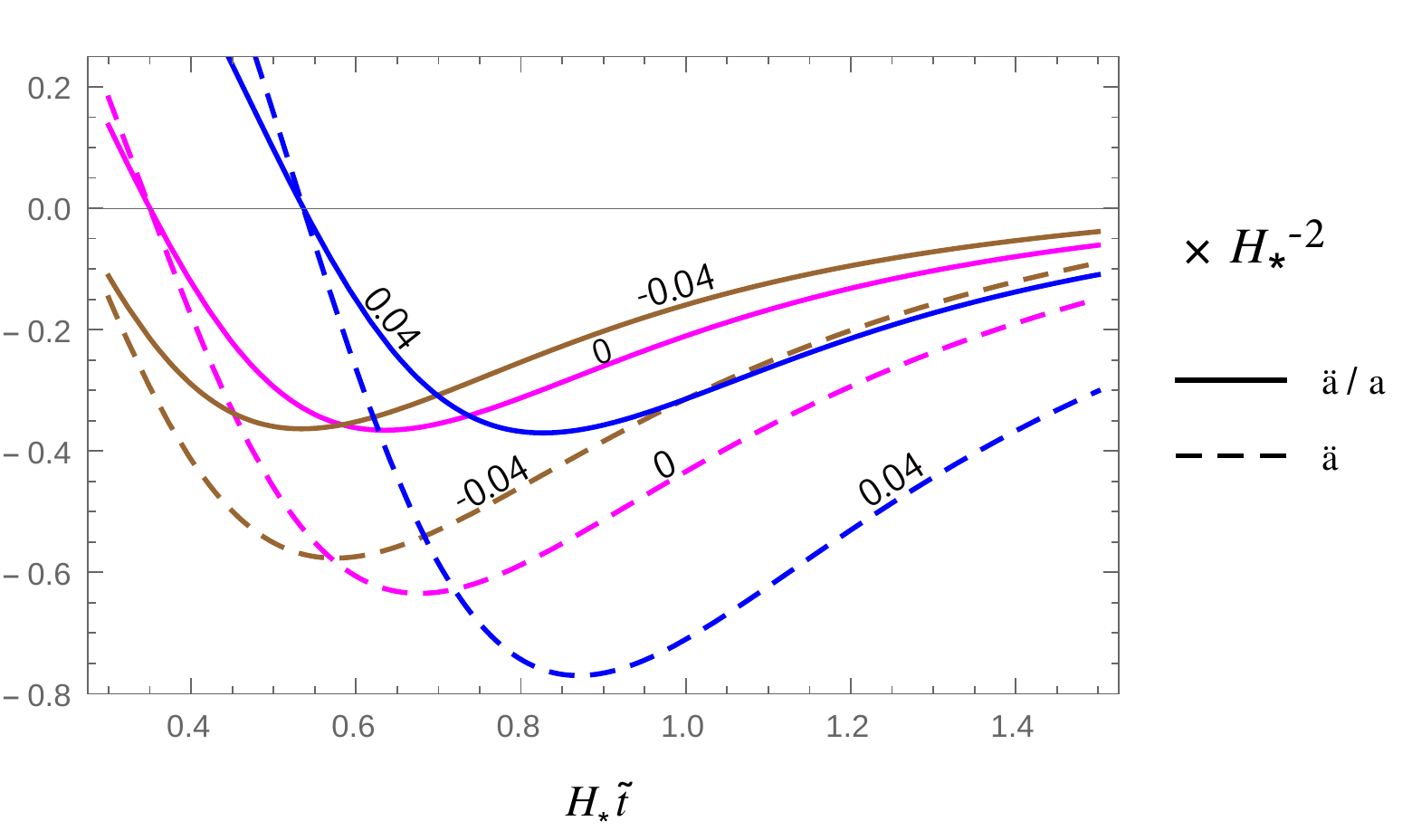}
\caption{\label{Analysis8}The temporal variation of $\ddot{a}/a$ and $\ddot{a}$ for a Universe comprising a VdW fluid and a dynamical $\Lambda \propto a^{-1/2}$. Each curve is labeled with the respective value of $\kappa$.}
\end{figure} 
\begin{figure}
\centering
\includegraphics[width=10cm]{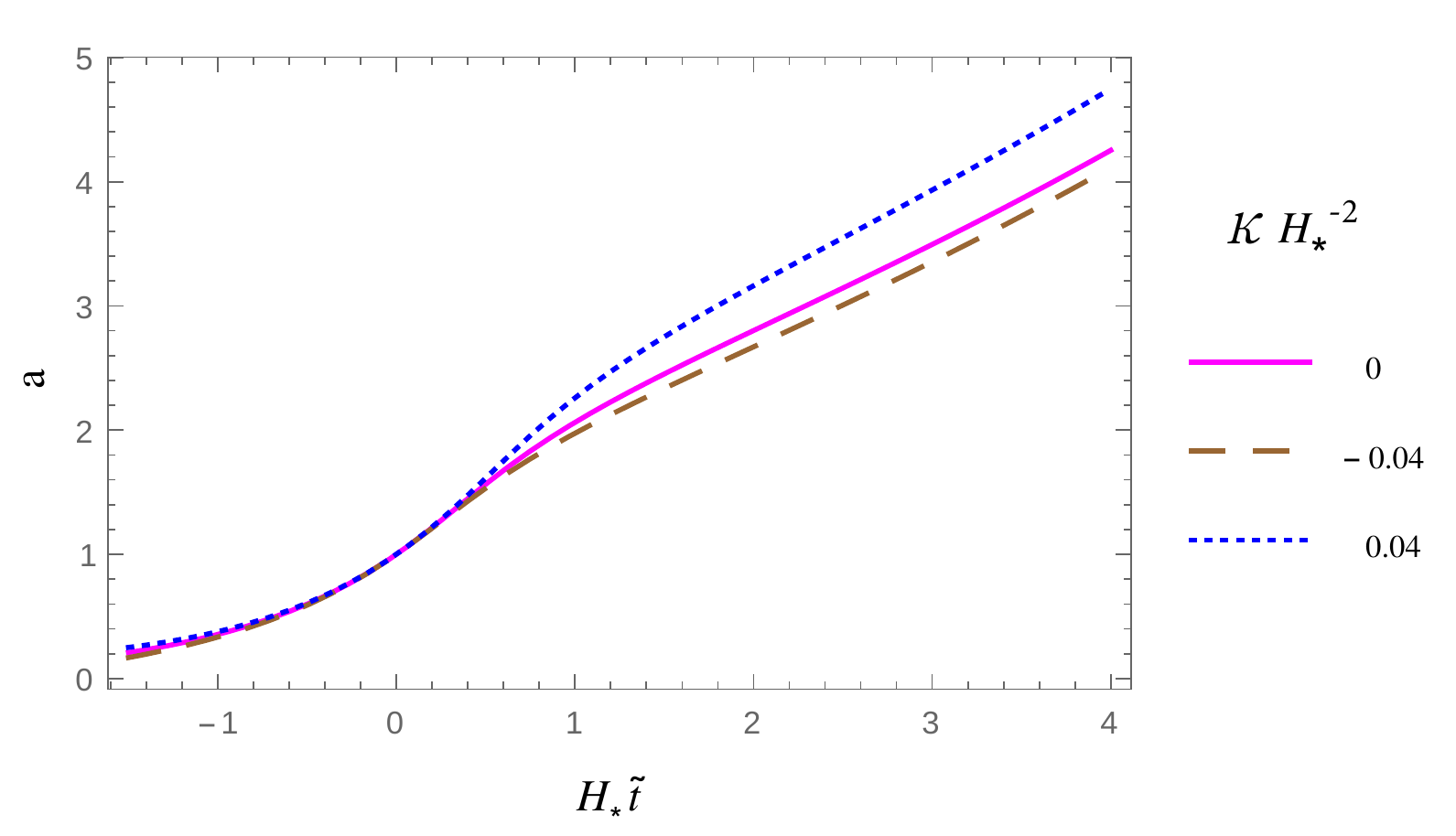}
\caption{\label{Lambda2a}The variation of the scale factor with time for a Universe consisting of a VdW fluid and a dynamical $\Lambda \propto a^{-1/2}$.}
\end{figure} 

The large repulsive pressure in the closed Universe is the main factor behind the delay in its transition from the first period of acceleration to the subsequent deceleration, as shown in Fig. $\ref{Analysis7}$. On the other hand, Fig. $\ref{Analysis8}$ demonstrates that the maximum deceleration differs for the flat, open and closed geometries due to the fact that the value of the scale factor varies with $\kappa$ (Fig. $\ref{Lambda2a}$). 

The dependence of $\ddot{a}$ on $a$ at any given time is expressed by the equation:
\begin{equation}
\ddot{a}=-\frac{a}{2}\left[\rho_\text{m}+3p_\text{m}-\frac{2}{3}\Lambda(\tilde{t})\right].
\label{a''2}
\end{equation} 

The terms in square brackets amount to the total sum of the energy densities and pressures (with the latter multiplied by three). As evident from Fig. $\ref{Analysis8}$, the said sum (equal to $-2\ddot{a}/a$) turns out to be comparable for all three values of $\kappa$ during the epoch of deceleration. Given the larger scale factor of the closed Universe (Fig. $\ref{Lambda2a}$), this essentially means that the total matter content must be greater for $\kappa>0$, which would then imply a resultant repulsion exceeding that for the flat and open Universes. We can thus explain why the amplitude of deceleration differs with $\kappa$.  

After some time, the pressure exerted by the VdW fluid becomes positive; in other words, it is now attractive. None the less, it remains largest in the closed\footnote{After $p_\text{m}$ becomes attractive, there is actually a short period of adjustment during which it is smallest for the closed Universe.} Universe, where the matter distribution is still the most dense. For late-time acceleration to set in, the repulsion produced by dark energy must outbalance the attraction due to matter. As shown in Fig. $\ref{Lambda2acc}$, this occurs first in the open Universe, where $\rho_\text{m}+3p_\text{m}$ is smallest.  

\section{Modeling Matter with EoS~ $\boldmath{p_\text{m}=w\rho_\text{m}}$}   
In this section, $t$ denotes ordinary cosmic time. We also revert to the usual definitions of $a$ and $\kappa$, so that the former is now equal to the scale factor when normalized with respect to its present value ($a=R/R_0$), and $\kappa = k/R_0^2$ (with $k=0$ or $\pm 1$). The matter distribution is described by the EoS:
\begin{equation}
p_\text{m}=w\rho_\text{m}
\label{Barotropic}
\end{equation}
where $w \in [0,1]$ is a parameter that depends on the type of fluid: $w=0$ represents dust, radiation has $w=1/3$, and a stiff fluid is one with $w=1$.

We combine matter with a dark energy component that evolves according to:\cite{Lima2015}
\begin{equation}
\Lambda(t)=\Lambda_\infty+3\left(\frac{H}{H_\text{I}}\right)^n\left(\frac{\kappa}{a^2}+H^2\right)
\label{Lambda3}
\end{equation} 
where $\Lambda_\infty$ is the value of $\Lambda(t)$ as $a\rightarrow \infty$, $H_\text{I}$ stands for the Hubble parameter associated with the primordial de Sitter stages, and $n\geq 1$ is an integer. At very early times, $H\sim H_\text{I}$ and the vacuum dynamics are driven by the second term on the right-hand side of ($\ref{Lambda3}$). When $H\ll H_\text{I}$, however, this term becomes negligible, rendering $\Lambda(t)$ approximately equal to $\Lambda_\infty$. Consequently, at late times the resulting cosmology converges to $\Lambda$CDM.\cite{Lima2015} 
\begin{figure}[b]
\centering
\includegraphics[width=10cm]{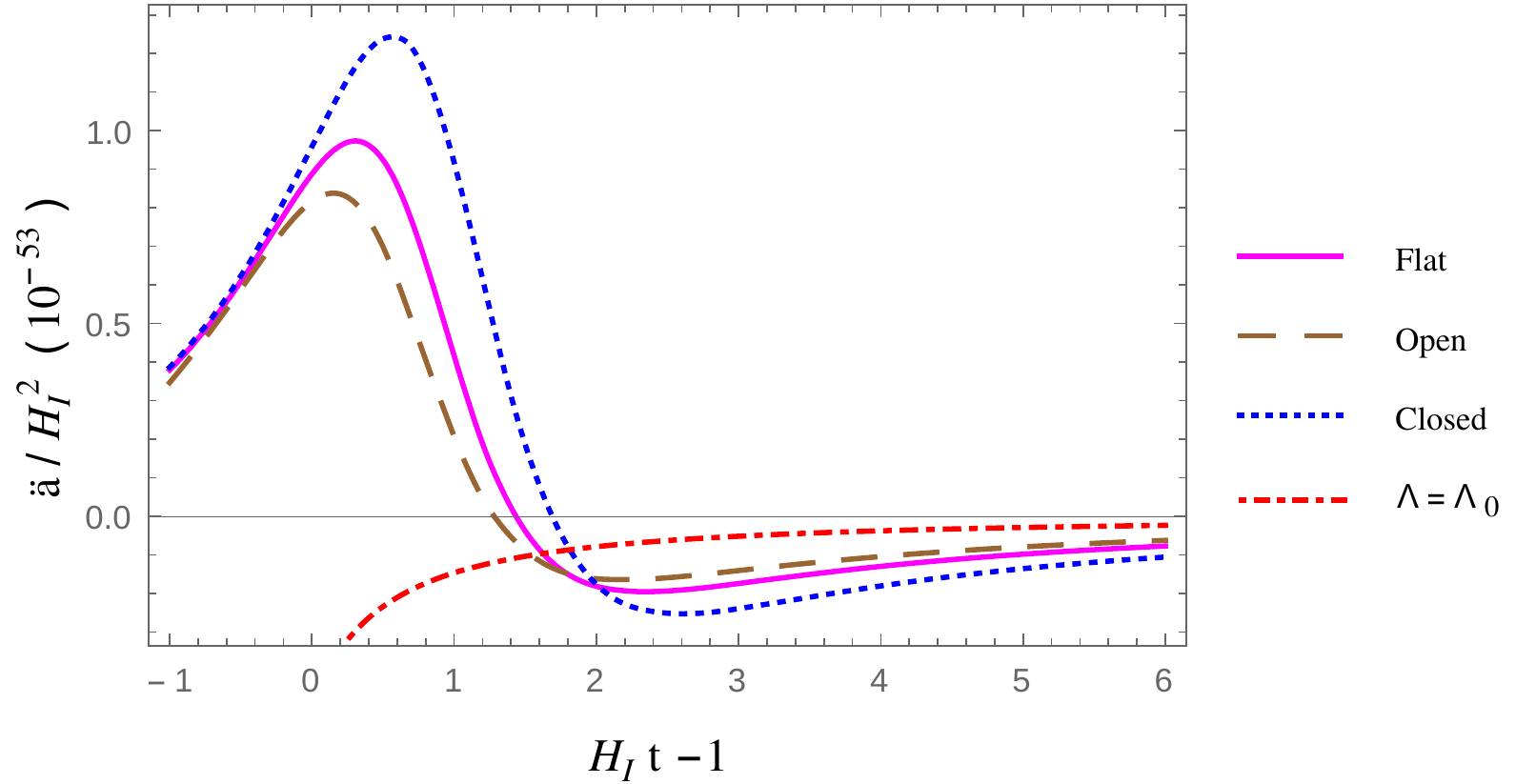}
\caption{\label{Lambda3acc}The variation of acceleration with time for a Universe composed of dust and a dynamical $\Lambda$ that evolves according to Eq. ($\ref{Lambda3}$). The acceleration of a \emph{flat} Universe, also containing dust but in which $\Lambda(t)$ is replaced by a cosmological constant, is represented by the (red) dot-dashed curve. In this case, no inflationary epoch precedes the period of decelerated expansion.}
\end{figure}
\begin{figure}[t]
\centering
\includegraphics[width=10cm]{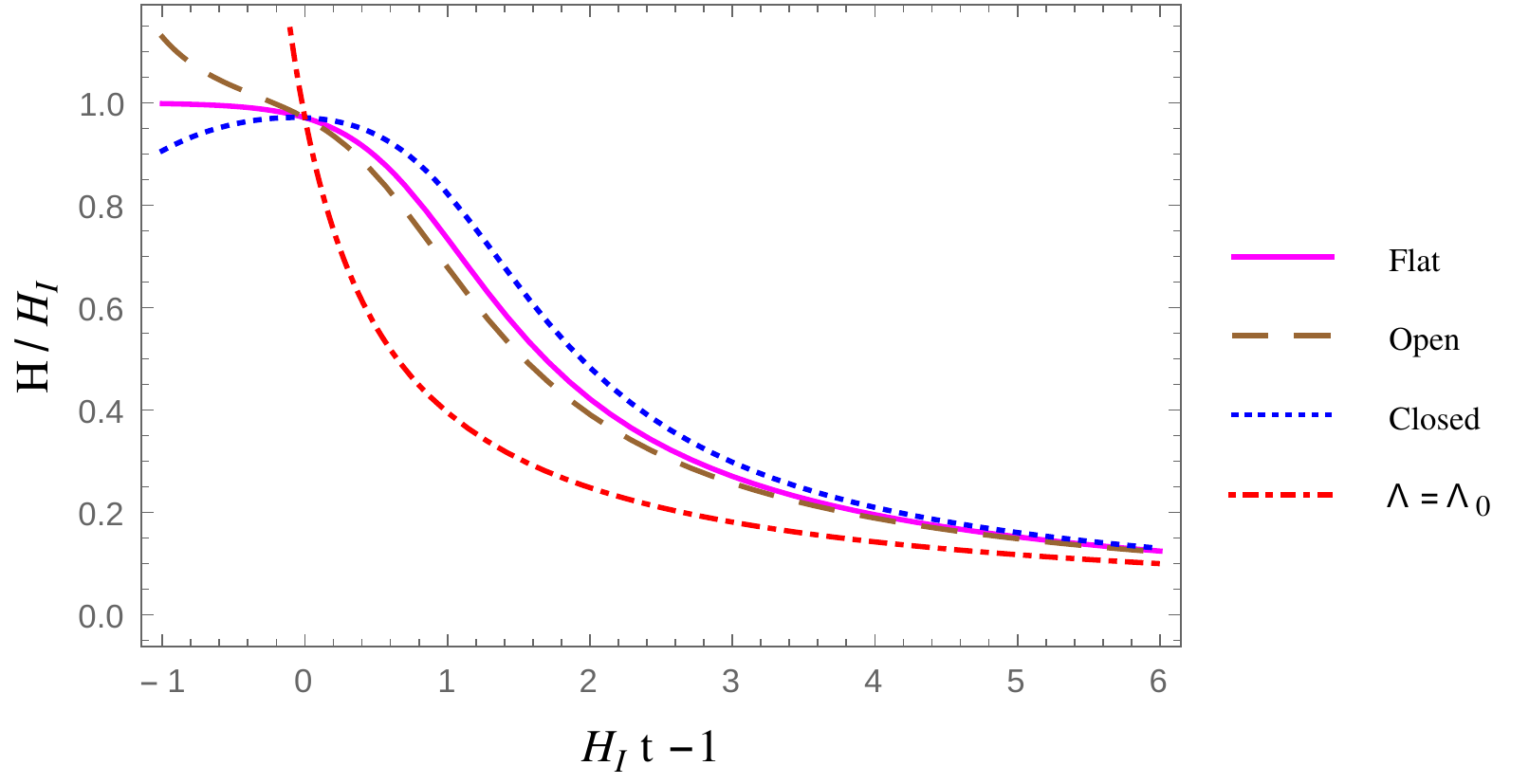}
\caption{\label{Lambda3H}The variation of the Hubble parameter with time for a Universe modeled as a composition of dust and a dynamical $\Lambda$ that evolves according to Eq. ($\ref{Lambda3}$). The Hubble parameter for a \emph{flat} Universe, also containing dust but in which $\Lambda(t)$ is replaced by a cosmological constant, is represented by the (red) dot-dashed curve.}
\end{figure}

Substituting ($\ref{Barotropic}$), together with $\rho_\text{m}$ as obtained from Eq. ($\ref{1stFriedmann3}$), into ($\ref{timederivative}$) yields the evolution equation for the system:
\begin{equation}
(\kappa+\dot{a}^2)\left\{3(1+w)\left[1-\left(\frac{\dot{a}}{H_\text{I}a}\right)^{n\phantom{!}}\right]-2\right\}-\Lambda_\infty(1+w)a^2+2a\ddot{a}=0
\label{MainEq5}
\end{equation}
where we have expressed $\Lambda(t)$ according to ($\ref{Lambda3}$). With a few modifications, Eq. ($\ref{MainEq5}$) is seen to be equivalent to Eq. (22) in\footnote{Care should be taken when comparing the two equations. In ($\ref{MainEq5}$), $a$ represents the normalized scale factor (rather than the \emph{non-normalized} version, as in Ref.~\refcite{Lima2015}), while $\kappa$ stands for the ratio $k/R_0^2$. The authors of Ref.~\refcite{Lima2015} use $\kappa$ to represent the \emph{normalized} curvature parameter.} Ref.~\refcite{Lima2015}. Since the authors point out that $\Lambda(t)$ attains its final value ($\Lambda_\infty$) at the start of the adiabatic radiation phase, we identify $\Lambda_{\infty}$ with $\Lambda_{0}$, the current value of $\Lambda(t)$.

We first consider the inflationary epoch -- together with its graceful exit to the period of cosmic deceleration -- and proceed by redefining the time coordinate as $\hat{t} = H_\text{I}t - 1$; $\hat{t}$ is thus dimensionless. With this definition, the Hubble parameter is automatically rescaled to $\hat{H}=H/H_\text{I}$. Hence, when $H=H_\text{I}$, $\hat{H}=1$ and, assuming the relation $t\sim 1/H$, one gets that $\hat{t}\approx 0$. The reference time ($\hat{t}=0$) is thus specified as the point at which the Hubble parameter is (approximately) equal to unity, in line with the procedure adopted in the previous section. Additionally, the rescaled time coordinate implies that Eq. ($\ref{MainEq5}$) can be rewritten in the form:
\begin{equation}
(\hat{\kappa}+\dot{\hat{a}}^2)\left\{3(1+w)\left[1-\left(\frac{\dot{\hat{a}}}{\hat{a}}\right)^{n\phantom{!}}\right]-2\right\}-3(1+w)\mathcal{H}^2\Omega_{\Lambda}^0\hat{a}^2+2\hat{a}\ddot{\hat{a}}=0
\label{MainEq5rescaled}
\end{equation}
where $\hat{\kappa}=\kappa/H_\text{I}^2$, $\dot{\hat{a}}=\dot{a}/H_\text{I}$, $\hat{a}=a$, $\ddot{\hat{a}}=\ddot{a}/H_\text{I}^2$, $\mathcal{H}=H_0/H_\text{I}$ and $H_0$ is the present value of the Hubble parameter. The relation $\Lambda_0=3H_0^2 \Omega_\Lambda^0$ was used to express $\Lambda_0$ in terms of the corresponding density parameter. 

We set the initial conditions to $\hat{a}(0) = 1.03\phantom{!}\mathcal{H}$ and $\dot{\hat{a}}(0) = \mathcal{H}$ (the dot now denotes differentiation with respect to $\hat{t}$\phantom{!}). The slight discrepancy between $\hat{a}(0)$ and $\dot{\hat{a}}(0)$, while being small enough to ensure that $\hat{H}(0)$ is still sufficiently close to unity, allows the numerical simulator to yield better results, because the first term enclosed in curly brackets in Eq. ($\ref{MainEq5rescaled}$) does not completely drop out of the equation at $\hat{t}=0$. 

The parameters are assigned values as follows: $|\hat{\kappa}| = \num{0.08}\phantom{!}\mathcal{H}^2$ (or zero), $w=0$ and $n=2$. The last tallies with the suggestion that $n$ should be even in flat space, as indicated by the covariance of the effective action of Quantum Field Theory in curved spacetimes.\cite{Perico, Lima2013} $\mathcal{H}$ is set equal to\footnote{This would mean that -- to a good approximation -- the beginning of the deflationary process coincides with grand unification \cite{Lima1996}. Our value for $\mathcal{H}$ differs by an order of magnitude from the one given in Ref.~\refcite{Lima1996}, however, because we use the value reported in Ref.~\refcite{Planck} for $H_0$. We also considered typical values for the Hubble parameter at the end of inflation\cite{Moreno}, as well as the current value\cite{Planck} of $H_0$, to calculate the magnitude of $\mathcal{H}$.} $\num{e-53}$, with $\Omega_\Lambda^0$ taking the value \num{0.692}. The latter was reported in Ref.~\refcite{Planck}, where the authors assume a $\Lambda$CDM cosmology. This, however, poses no serious issue, since our model should converge to $\Lambda$CDM at late times. 

\begin{figure}[t]
\centering
\includegraphics[width=10cm]{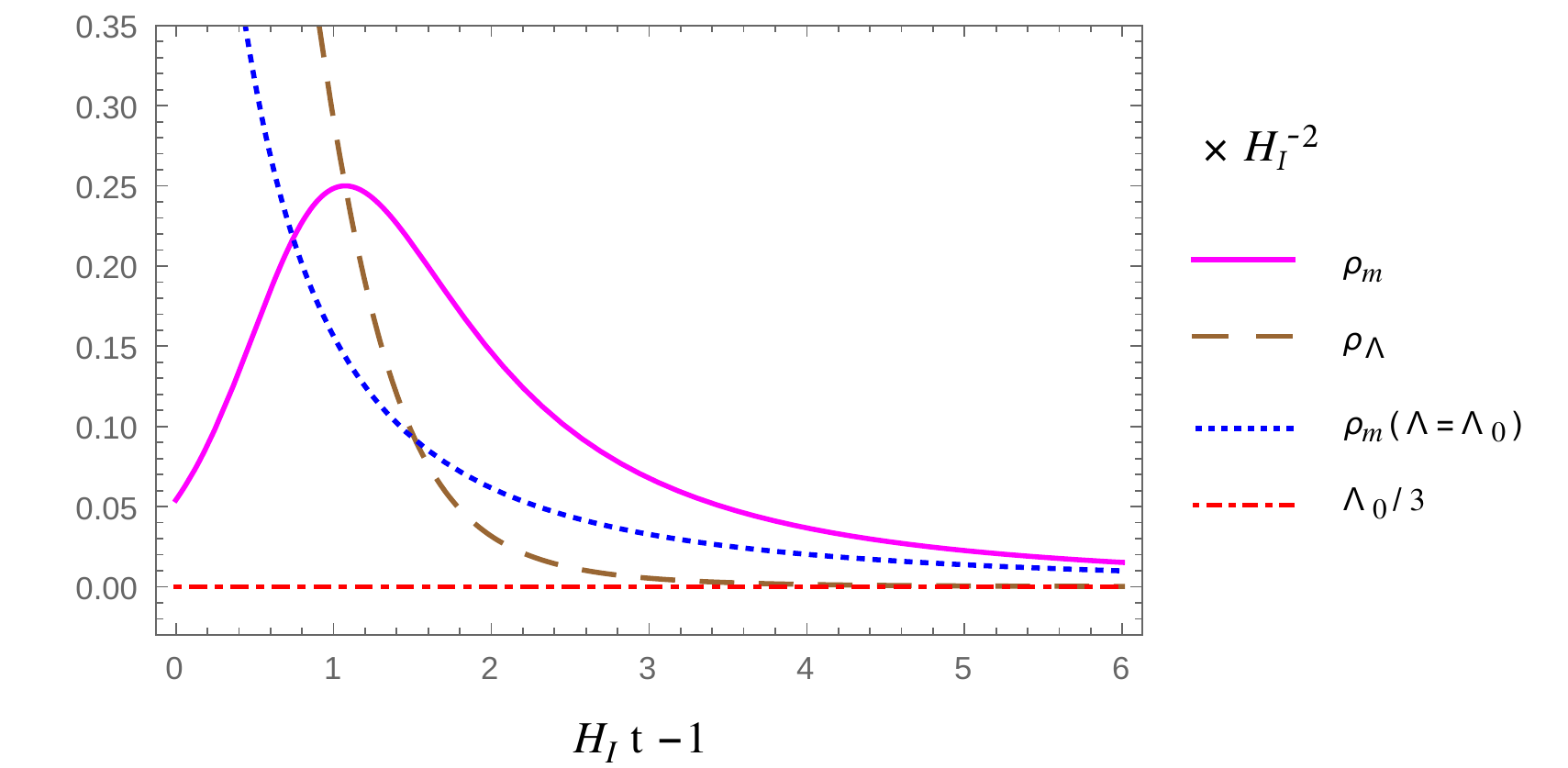}
\caption{\label{Lambda3rho1}The temporal variation of $\rho_\text{m}$ and $\rho_\Lambda$ at the end of inflation and the beginning of the matter-dominated epoch. Only a flat spatial geometry is considered. The matter component is modeled as dust, with dark energy described by a dynamical $\Lambda$ that takes the form specified in Eq. ($\ref{Lambda3}$). The evolution of $\rho_\text{m}$ in the presence of a cosmological constant is also depicted. The energy density associated with this constant is represented by the (red) dot-dashed curve.}
\end{figure}

Figures $\ref{Lambda3acc}$ and $\ref{Lambda3H}$ show the temporal evolution of $\ddot{\hat{a}}$ and $\hat{H}$ during the inflationary epoch and the first stages of cosmic deceleration. At early times, the second term on the right-hand side of ($\ref{Lambda3}$) determines the behavior of $\Lambda(t)$ -- and in turn, $\Lambda(t)$ controls the cosmic evolution, as evidenced by Fig. $\ref{Lambda3rho1}$. Since the term in question is a function of $\kappa$, the history of the early Universe depends significantly on the spatial geometry. It can be seen that the effects of curvature match those observed for all the other models.

We also include the evolution according to the flat $\Lambda$CDM cosmology for purposes of comparison. This is based on the equation:
\begin{equation}
2\hat{a}\ddot{\hat{a}}+\dot{\hat{a}}^2+\mathcal{H}^2(\Omega_\text{r}^0\hat{a}^{-2}-3\Omega_\Lambda^0\hat{a}^2)=0.
\label{LCDM}
\end{equation}
$\Omega_\text{r}^0$ is the present-day value of the radiation density parameter, equivalent to\footnote{This can be calculated from the current value of the matter density parameter,\cite{Planck} $\Omega_\text{m}^0=\num{0.308}$, and the redshift at which the matter and radiation densities were equal,\cite{Planck} $z_\text{eq}=\num{3365}$.} $\num{9.15e-5}$. Its smallness implies that any effects of radiation on the dynamics of the Universe are negligible at late times. 

In the $\Lambda$CDM cosmology, the Universe is modeled as a mixture of dust, radiation and dark energy, with the last taking the form of a cosmological constant $\Lambda$. The dust component consists of both baryonic and dark matter, and its density varies as $a^{-3}$. Radiation density is proportional to $a^{-4}$ and thus gets diluted faster as the Universe expands. On the other hand, the density of dark energy is not affected by cosmic dynamics and remains constant throughout the entire evolution. Cosmological observations put the current values of the matter and dark energy density parameters at $\Omega_\text{m}^0\approx\num{0.3}$ and $\Omega_\Lambda^0\approx\num{0.7}$, respectively, which implies that we are living in an epoch dominated by dark energy. On the other hand, the radiation density parameter is at present negligibly small. At early times, however, radiation would have been very significant. Indeed, it is thought that the matter-dominated epoch was preceded by a period of time during which radiation was the dominant component of the cosmic fluid. Nonetheless, our model is compared with a $\Lambda$CDM cosmology consisting solely of dust and the cosmological constant. We are thus better able to investigate the effects of a dynamical $\Lambda$, this being the only difference between the two models. Strictly speaking, therefore, our results are compared with the ones predicted by a constant-$\Lambda$ model, rather than by $\Lambda$CDM. This is especially true at early times.

Fig. $\ref{Lambda3acc}$ shows that the maximum deceleration attained during the matter-dominated epoch is smaller when dark energy is modeled as a constant $\Lambda$ than when it is allowed to evolve according to Eq. ($\ref{Lambda3}$). This is due to the fact that in the latter case, the vacuum transfers energy to the matter component as it decays (refer to Eq. ($\ref{energycons4}$) and Fig. $\ref{Lambda3rho1}$), causing its energy density to grow. The deceleration resulting from the attractive pressure of matter is consequently larger than in the constant-$\Lambda$ scenario, where such a mechanism is absent. 

As pointed out in Ref.~\refcite{Lima2015} and shown in Fig. $\ref{Lambda3H}$, only a flat Universe would initially have a constant Hubble parameter: the effect of positive curvature would be to decrease the initial $\hat{H}$ below that in the flat case, while for negative $\kappa$ the plot of $H/H_\text{I}~(=\hat{H})$ against $\hat{t}$ should be truncated at $H/H_\text{I}=1$ (otherwise the weak energy condition would be violated). The early evolution of the Hubble parameter in the previous model (Fig. $\ref{Lambda2H}$) has similar features: $H/H_*$ is initially (approximately) constant in a flat Universe, while positive curvature makes it smaller. A hyperbolic geometry has the opposite effect. These characteristics also emerge in the model consisting solely of a VdW fluid (Fig. $\ref{VdWH}$), and are consistent with the properties of the primordial de Sitter solutions outlined in Ref.~\refcite{Lima2015}. 

\begin{figure}[b]
\centering
\includegraphics[width=10cm]{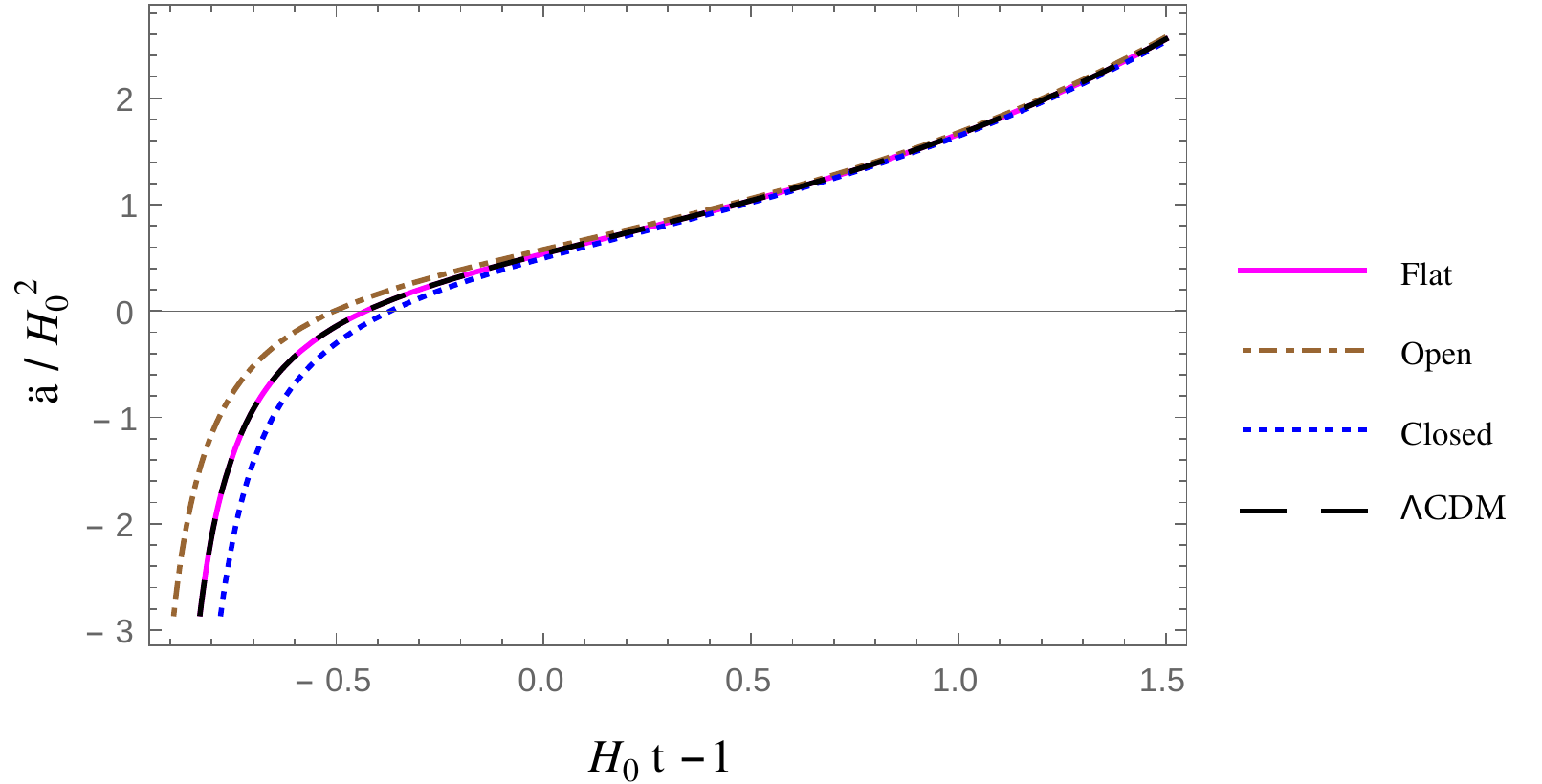}
\caption{The variation of acceleration with time during the current period of cosmic acceleration and the latter part of the preceding matter-dominated epoch. The Universe is modeled as a composition of dust and a dynamical $\Lambda$ that evolves according to Eq. ($\ref{Lambda3}$). At this stage, the acceleration of the flat Universe is indistinguishable from the one predicted by the flat $\Lambda$CDM cosmological model.}
\label{Lambda3late1}
\end{figure} 

\begin{figure}[b]
\centering
\includegraphics[width=10cm]{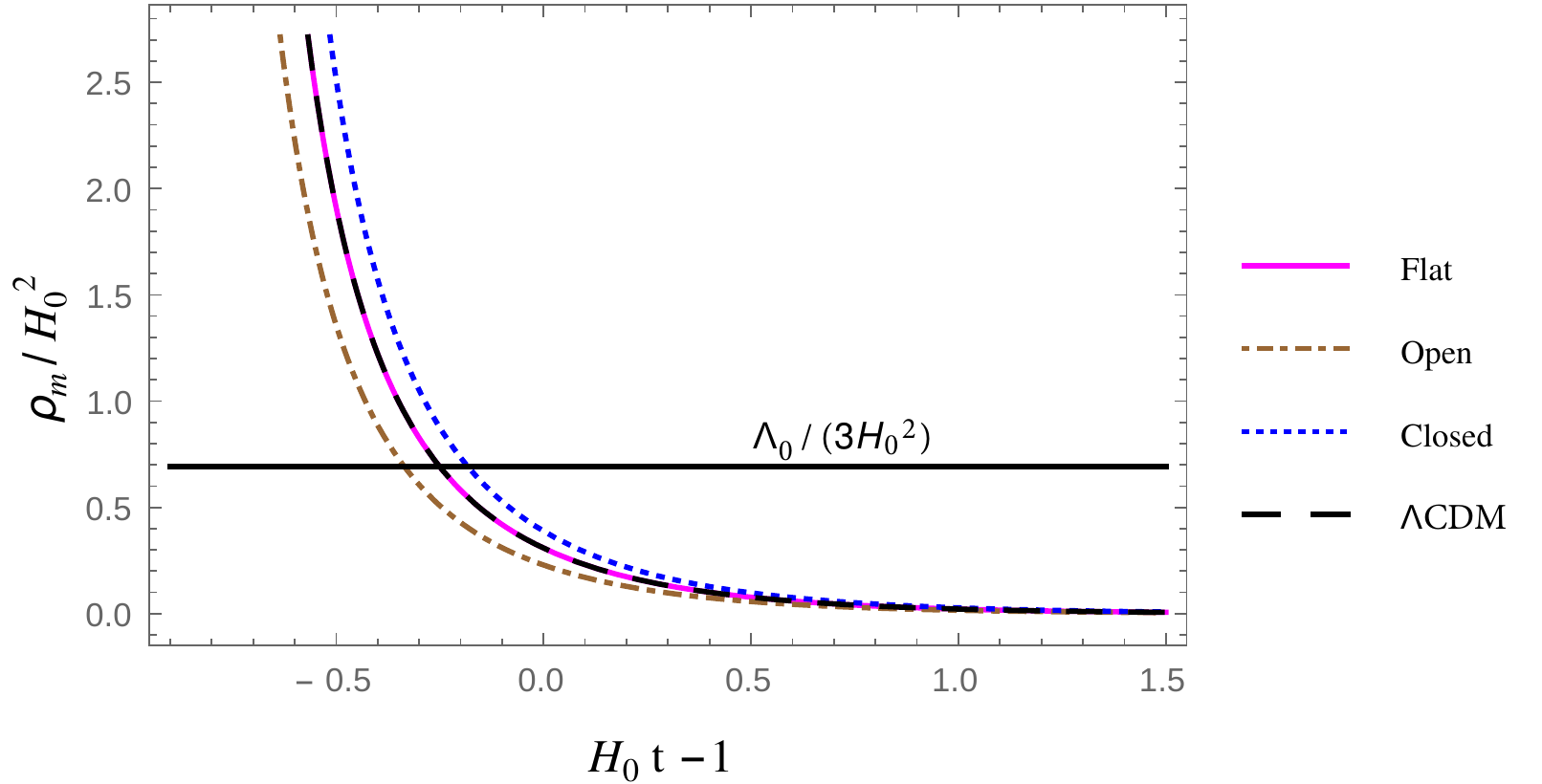}
\caption{\label{Lambda3rho2}The variation of $\rho_\text{m}$ with time during the transition from the matter-dominated epoch to the current period of cosmic acceleration. The Universe is modeled as a composition of dust and a dynamical $\Lambda$ that evolves according to Eq. ($\ref{Lambda3}$). The dashed curve indicates the energy density of dust in a flat $\Lambda$CDM cosmology, which would vary according to $\rho_\text{m}=\Omega_\text{m}^0 H_0^2 a^{-3}$. $\Lambda(t)$ would already have started to behave as a cosmological constant ($\Lambda_0$) at this stage. The energy density associated with $\Lambda_0$ does not change with time and is depicted by the straight horizontal line.}
\end{figure} 

We now turn to investigating the evolution of the Universe at late times. For this purpose, the time parameter is redefined as $\bar{t}=H_0 t-1$. Hence, if one assumes the relation $t\sim 1/H$, the current time corresponds to $\bar{t}=0$. The Hubble parameter, which is automatically rescaled to $\bar{H}=H/H_0$, becomes equal to unity at present. 

Eq. ($\ref{MainEq5}$) can be rewritten as:
\begin{equation}
(\bar{\kappa}+\dot{\bar{a}}^2)\left\{3(1+w)\left[1-\left(\frac{\dot{\bar{a}}}{\bar{a}}\mathcal{H}\right)^{n\phantom{!}}\right]-2\right\}-3(1+w)\Omega_\Lambda^0\bar{a}^2+2\bar{a}\ddot{\bar{a}}=0
\end{equation}
while Eq. ($\ref{LCDM}$) takes the form:
\begin{equation}
2\bar{a}\ddot{\bar{a}}+\dot{\bar{a}}^2+\Omega_\text{r}^0\bar{a}^{-2}-3\Omega_\Lambda^0\bar{a}^2=0
\end{equation}
where $\bar{\kappa}=\kappa/H_0^2$, $\dot{\bar{a}}=\dot{a}/H_0$, $\bar{a}=a$, $\ddot{\bar{a}}=\ddot{a}/H_0^2$, and the dot denotes differentiation with respect to $\bar{t}$. The initial conditions are set to $\bar{a}(0)=\dot{\bar{a}}(0)=1$. The first is required by the definition of the normalized scale factor, and the second follows from the fact that $\bar{H}(0)=1$. The parameters $\bar{\kappa}$, $n$ and $w$ are assigned the same values as for the early evolution. Since $\bar{\kappa}=\mathcal{H}^{-2}\hat{\kappa}$, however, we have that $|\bar{\kappa}|$ is equal to \num{0.08} (or zero if the Universe is flat).

The resulting late-time dynamics are depicted in Fig. $\ref{Lambda3late1}$. It can again be noted that the open Universe would be the first to exit the matter-dominated epoch. As expected, the evolution according to the flat $\Lambda$CDM cosmology (in which radiation can essentially be considered absent at this stage) would have become indistinguishable from that of the flat, dynamical-$\Lambda$ scenario by this time. Additionally, the presence of curvature no longer affects the evolution significantly after the transition to the current epoch of acceleration. This is due to the fact that, as shown in Fig. $\ref{Lambda3rho1}$, the energy of the vacuum decays rapidly and soon starts to behave as a cosmological constant, whose value is the same for all $\kappa$. Consequently, when at late times dark energy again starts to dominate (see Fig. $\ref{Lambda3rho2}$), its lack of dependence on the spatial geometry results in an evolution that is common to the flat, closed and open Universes. 

\section{Conclusion}
Many works in the literature are premised on the assumption of spatial flatness. However, the majority of the recent studies which conclude that observational data is consistent with a flat geometry fit the said data to the $\Lambda$CDM model, or extensions thereof.

The question of whether the Universe is actually flat is not only of interest in itself. As pointed out in Refs.~\refcite{Clarkson} and \refcite{Hlozek}, the assumption of zero spatial curvature -- if wrong -- could greatly undermine the efforts of the scientific community to construct an EoS for dark energy, even if the curvature is in reality only very small. Given that the nature of dark energy is at present one of the most important problems in cosmology, the presence or absence of cosmic curvature is an issue which merits our attention, especially since a number of studies have shown that fitting observational data to models with a time-varying dark energy EoS can, in some cases, actually accommodate a non-flat Universe.\cite{Ichikawa, IchikawaK}

In this work, we depart from the popular assumption of spatial flatness and investigate the effects of curvature on cosmic evolution by constructing five toy models. We find that for a closed Universe, the transition to the epoch of decelerated expansion would be delayed with respect to the flat case. So would the start of the current dark energy-dominated era. This would be accompanied by a larger inflationary acceleration, as well as a larger subsequent deceleration. On the other hand, the opposite would happen for an open Universe. The fact that these characteristics are common to all five models\footnote{Or to the last four, in the case of late-time acceleration.} implies that the changes introduced by the presence of curvature are independent of the way the matter and dark energy components are modeled. We also note that small variations in the initial conditions or the parameters only influence the magnitude of the above-mentioned effects, and not the behavior of the evolution. To a good approximation, it can be said that in all models, the cosmic evolution is independent of the spatial geometry at late times. 
 
For the first model, we endow the cosmic fluid with the VdW EoS and confirm that a dark energy component is needed to correctly reproduce the current period of acceleration. Thus, in the remaining four models we introduce dark energy as Quintessence, a Chaplygin gas or a dynamical cosmological `constant', and get the desired late-time acceleration. The matter distribution is represented by a VdW fluid in the first four models, with the EoS being changed to the customary $p_\text{m}=w \rho_\text{m}$ for the last one. We follow Refs.~\refcite{Jafarizadeh} and \refcite{Lima2015} when dealing with dark energy as a dynamical $\Lambda$, and the work of Kremer\cite{Kremer} elsewhere.  

A unique characteristic of the last model -- apart from the proportionality between the energy density and pressure of the matter component -- is a time-dependent vacuum energy that replaces the inflaton at early times, and reproduces the effects of the cosmological constant during the present epoch.\cite{Lima2015} We compare our results with those for a flat Universe made up of dust and a cosmological constant (and hence equivalent to the flat $\Lambda$CDM scenario at late times), and find that the deceleration during the matter-dominated epoch would be greater when the energy of the vacuum is allowed to vary with time. This is due to the fact that as $\Lambda(t)$ decays, energy is transferred to the matter component, increasing its density and causing it to further slow down the cosmic expansion. 

In conclusion, a non-zero $\Omega_k$ would have left a definite signature on the past dynamics of the Universe. This means that it is possible to look for the potential presence of curvature by analyzing observational data for a wider variety of dark energy models or, better still, independently of any at all.

\section*{Acknowledgements}
The research work presented in this paper is partially funded by the Endeavour Scholarships Scheme. The scholarship may be part-financed by the European Union -- European Social Fund.

\bibliographystyle{ws-ijmpd}
\bibliography{reference}

\end{document}